\documentclass[aps,pre,preprint,groupedaddress,showpacs]{revtex4-1}
\usepackage[dvips]{graphicx}
\usepackage{textcomp}
\usepackage{amssymb}

\newcommand{\mc}{\multicolumn}

\begin{document}

\title{
\Large\bf Eliminating leading and subleading corrections to scaling in the 
three-dimensional XY universality class}

\author{Martin Hasenbusch}
\affiliation{
Institut f\"ur Theoretische Physik, Universit\"at Heidelberg,
Philosophenweg 19, 69120 Heidelberg, Germany}

\date{\today}
\begin{abstract}
We study the $(q+1)$-state clock model on the simple cubic lattice
by using Monte Carlo simulations.
In addition to the nearest neighbor coupling we consider a 
next-to-next-to-nearest neighbor coupling. 
For a certain range of the parameters, the phase transition of
the model shares the XY universality class. 
Leading corrections to scaling
are studied by using finite size scaling of dimensionless quantities, 
such as the Binder cumulant $U_4$. The spatial unisotropy, which causes
subleading corrections, is studied 
by computing the exponential correlation length $\xi_{exp}$ in the high 
temperature phase for different directions. In the case of the 
$q$-state clock model it turns out that by
tuning the ratio of the two coupling constants, we can eliminate either 
leading or subleading corrections to scaling. These points on the critical
line are close to each other. Hence in the improved model, where 
leading corrections to scaling vanish, also subleading corrections
are small.
By using a finite size scaling analysis of our high statistics data we 
obtain $\eta=0.03816(2)$ and $y_t =1/\nu=1.48872(5)$ as estimates 
of the critical exponents. 
\end{abstract}

\maketitle
\section{Introduction}
We discuss the extension of the improved model programme, studying
critical phenomena, to a two-parameter family of models at the example of the
three-dimensional XY universality class. For reviews on critical phenomena 
see for example \cite{WiKo,Fisher74,Cardy,Fisher98,PeVi02}.
For a discussion of improved lattice models see in particular Sec. 2.3 of
Ref. \cite{PeVi02}.
Studying critical phenomena by using Monte Carlo simulations of lattice models
one might have different objectives.
Either the model reproduces well the microscopic properties of
the experimental system of concern or one is interested in properties of 
a universality class. In the latter case, we would like to obtain accurate 
estimates of universal quantities like critical exponents or amplitude ratios 
by simulating relatively small lattices, allowing for high statistics.  
To this end, the model should not be too complicated and one should be able 
to simulated it by using efficient algorithms such as the cluster
algorithm \cite{SwWa87,Wolff}.
Here we study a model in the three-dimensional XY universality class, which
fulfills these criteria. In Ref. \cite{myIso} we studied the Blume-Capel 
model on the simple cubic lattice with nearest and next-to-next-to-nearest
neighbor couplings. By tuning the ratio of the two coupling constants 
and the dynamic dilution parameter, also called crystal field, we were able to 
find a point in the parameter space, on the critical surface, where both the 
leading correction as well as the subleading correction, which is 
associated with the spatial unisotropy, are vanishing.
In the case of the Ising universality class, the corresponding 
correction exponents are $\omega=0.82968(23)$, and 
$\omega_{NR}=2.022665(28)$, obtained by the using the conformal bootstrap (CB) 
method \cite{Kos16,SD16}. Note that these estimates are consistent with
previous, less accurate estimates obtained by using Monte Carlo simulations and
series expansions of lattice models or field theoretic methods.
The finite size scaling (FSS) analysis of the Blume-Capel with nearest and
next-to-next-to-nearest neighbor couplings results in the most accurate 
estimates
of critical exponents by using Monte Carlo simulations of lattice models
for the three-dimensional Ising universality class. The accuracy is clearly
better than that of field-theoretic methods. It is 
only surpassed by the CB \cite{Kos16,SD16,Re22,CB_Ising_2024}.   

Here we study an analogous model with phase transitions in the XY universality
class. To this end, the field variable $s_x \in \{-1,0,1\}$ is replaced by a 
two component vector, which has either unit length or vanishes. The precise
definition of the model and a discussion of its phase diagram will be given 
below in sec. \ref{themodel}. As already 
anticipated in the appendix of Ref. \cite{myIso}, where preliminary 
work is reported, we do not find a point, where the amplitudes of 
both corrections exactly vanish. However there is a point in parameter space, 
where the amplitude of the leading correction vanishes and the spatial 
unisotropy is considerably reduced. This allows us to further improve 
the accuracy of critical exponents compared with Ref. \cite{myClock}. 
Our estimates are fully compatible with those obtained by using the 
CB \cite{che19} and the error bars are of similar size.
For recent reviews on the CB see for example \cite{PaRyVi18,RySu23}.

The three-dimensional XY universality class has attracted much attention,
since the $\lambda$-transition of $^4$He, which is well studied 
experimentally, is supposed to share this universality class.
The most accurate result for the exponent $\alpha$ of the specific heat
is obtained from an experiment under the condition of microgravity
\cite{Lipa96,Lipa00,Lipa03}: $\alpha = - 0.0127(3)$, which corresponds 
to $\nu = (2-\alpha)/d = 0.6709(1)$, exploiting the hyperscaling relation
between the critical exponent of the correlation length $\nu$ and $\alpha$.
Field theoretic results at the time were 
considerably less accurate. See for example Ref. \cite{GuZi98}, where
$\nu=0.6703(15)$ and $0.6305(25)$ are obtained by using the $d=3$ expansion 
and the $\epsilon$-expansion, respectively.

Monte Carlo simulations in combination
with the analysis of high temperature series of lattice models
resulted in $\nu=0.6717(1)$ \cite{XY2}, reaching the accuracy claimed by
the experiment. This estimate was confirmed by more recent Monte Carlo
simulations of lattice models, $\nu=0.67183(18)$ \cite{Xu19} and 
$\nu=0.67169(7)$ \cite{myClock}, and the estimate $\nu=0.671754(99)$ 
obtained by using the CB \cite{che19}. There can be little doubt
on the theoretical estimate of $\nu$ now, and it would be desirable to 
reanalyze the experimental results or even perform new ones.
Our motivation for the present work is to further develop the concept
of improved models and to provide accurate estimates of critical exponents
as benchmark for other theoretical methods. 

The outline of the paper is the following: In Sec. \ref{themodel} we
define the model and introduce the observables
that are measured. In Sec. \ref{Corrections_summary} we summarize
results on corrections to scaling exponents given in the literature.
Next in Sec. \ref{Restoring} we analyze our data for the correlation length
in the high temperature phase and compute the optimal ratio of the 
coupling constants. In Sec. \ref{FSSnumerics}  follows a  FSS
analysis of dimensionless quantities. We obtain accurate estimates of critical
exponents.
In Sec. \ref{summary} we summarize our results and compare them with 
the literature. 

\section{The model}
\label{themodel} 
We consider the analogue of the Blume-Capel model
with nearest neighbor and next-to-next-to-nearest neighbor interactions
studied in Ref. \cite{myIso}. We replace
$\mathbb{Z}_2$ by $\mathbb{Z}_q$ or, in the limit $q \rightarrow \infty$, by
$O(2)$ symmetry.
The model studied in Ref. \cite{myIso} is defined by the reduced Hamiltonian
\begin{equation}
\label{BlumeCapel}
H = -K_1 \sum_{<xy>}  s_x s_y - K_3 \sum_{[xy]}  s_x s_y
  + D \sum_x s_x^2  -h \sum_x s_x  \;\; ,
\end{equation}
where the spin or field variable $s_x$ might assume the values 
$s_x \in \{-1, 0, 1 \}$ and
$x=(x^{(0)},x^{(1)},x^{(2)})$ denotes a site on the simple cubic lattice,
where $x^{(i)} \in \{0,1,...,L_i-1\}$. Furthermore, $<xy>$ denotes a pair 
of nearest and $[xy]$ a pair of next-to-next-to-nearest, or third nearest (3N)
neighbors on the lattice. In Ref. \cite{myIso} 
 the linear lattice size $L=L_0=L_1=L_2$ is equal in all three directions
and periodic boundary conditions are employed. 

The model defined below is more general than the one actually 
simulated and includes the model studied in Ref. \cite{myClock} as 
special case.
Here we are interested in the properties of the $O(2)$-invariant 
fixed point. 
Finite values of $q$ are taken for technical reasons: The simulation
requires less CPU time and less memory.  
Note that for example in Refs. \cite{LiTe89,KoOk14} the $512$-state clock  
model had been simulated as extremely good approximation of the XY model at 
the critical point. The reduced Hamiltonian is given by
\begin{equation}
\label{Hamilton}
 {\cal H}_{q+1} = -  K_1 \sum_{\left<xy\right>}  \vec{s}_x \cdot
     \vec{s}_y 
  -  K_3 \sum_{[xy]} \vec{s}_x \cdot \vec{s}_y 
-D  \sum_x \vec{s}_x^{\,2} - \vec{H} \cdot \sum_x \vec{s}_x \;.
\end{equation}
The geometry of the lattice is the same as above for the Blume-Capel model.
Whereas the field takes the values
\begin{equation}
\vec{s}_x \in
\left\{(0,0), \left(\cos(2 \pi m/q), \sin(2 \pi m/q)  \right)
\right\} \;,
\end{equation}
where $m \in \{1,...,q\}$. Compared with the $q$-state clock model,
$(0,0)$ is added as possible value of the field variable. Here we perform 
simulations for a vanishing external field $\vec{H}=\vec{0}$. 
Note that, following the convention of, for example \cite{XY1,myClock},
the parameter $D$ has the opposite sign as for the Blume-Capel model, 
Eq.~(\ref{BlumeCapel}). 
In the simulation program we use a label $m \in \{0,...,q\}$  to store 
the spins:
\begin{equation}
 \vec{s}(0) = (0,0)
\end{equation}
and for $m>0$
\begin{equation}
\label{directions}
 \vec{s}(m) =  \left(\cos(2 \pi m/q), \sin(2 \pi m/q)   \right) \;\;.
\end{equation}

We introduce the weight factor
\begin{equation}
\label{weight}
w(\vec{s}_x)  = \delta_{0,\vec{s}_x^{\,2}} +  
\frac{1}{q} \delta_{1,\vec{s}_x^{\,2}}
              = \delta_{0,m_x} + \frac{1}{q} \sum_{n=1}^q \delta_{n,m_x}
\end{equation}
that gives equal weight to $(0,0)$ and the collection of all values 
$|\vec{s}_x| =1$. Now the partition function can be written as
\begin{equation}
 Z = \sum_{\{\vec{s}\} }  \prod_x w(\vec{s}_x) \; \exp(-{\cal H}_{q+1}) \;,
\end{equation}
where $\{\vec{s}\}$ denotes a configuration of the field. The weight,
Eq.~(\ref{weight}), is introduced to get the measure of the dynamically 
diluted XY (ddXY) model \cite{XY1,XY2}
\begin{equation}
\label{measure}
d\mu(\vec{s}_x) =  d s_x^{(1)} \, d s_x^{(2)} \,
\left[
\delta(\phi_x^{(1)}) \, \delta(\phi_x^{(2)})
 + \frac{1}{2 \pi} \, \delta(1-|\vec{\phi}_x|)
\right] 
\end{equation}
in  limit $q \rightarrow \infty$.

For reasons that become clear below, we do not simulate at general $(D,K_3)$,
but only the two limiting cases $K_3 =0$ or $D \rightarrow \infty$.
For $K_3=0$ we get the model studied in Ref. \cite{myClock}. Here we 
perform new simulations for $q=12$ at criticality. Our main focus is however
on simulations for $D \rightarrow \infty$, where the relative
weight of $\vec{s}_x=(0,0)$ vanishes. Hence we simulate the $q$-state clock 
model with nearest and next-to-next-to-nearest neighbor interactions defined
by the reduced Hamiltonian
\begin{equation}
\label{Hamilton_Clock}
 {\cal H}_{q} = -  K_1 \sum_{\left<xy\right>}  \vec{s}_x \cdot
     \vec{s}_y
  -  K_3 \sum_{[xy]} \vec{s}_x \cdot \vec{s}_y
 - \vec{H} \cdot \sum_x \vec{s}_x \;,
\end{equation}
where the field takes the values
\begin{equation}
\vec{s}_x \in \left \{
 \left(\cos(2 \pi m/q), \sin(2 \pi m/q)  \right)
\right\} \;,
\end{equation}
where $m \in \{1,...,q\}$. All $q$ possible values of the spin 
have the same weight. Below, we denote this model by 3N $q$-state 
clock model. 

\subsection{Definitions of the measured quantities}
\label{def}
The quantities studied in FSS are 
the same as in \cite{myClock}. For completeness we list them below:
The energy density is defined as
\begin{equation}
\label{energy}
E=\frac{1}{V} \sum_{\left<xy\right>} \vec{s}_x \cdot \vec{s}_y\; .
\end{equation}
The magnetic susceptibility $\chi$ for a vanishing magnetization
and the second moment correlation length $\xi_{2nd}$ are defined as
\begin{equation}
\label{chidef}
\chi  =  \frac{1}{V} \,
\biggl\langle \Big(\sum_x \vec{s}_x \Big)^2 \biggr\rangle
\end{equation}
and
\begin{equation}
\xi_{2nd}  =  \sqrt{\frac{\chi/F-1}{4 \sin^2 \pi/L}} \;,
\label{xidef}
\end{equation}
where
\begin{equation}
F  =  \frac{1}{V} \, \biggl\langle
\Big|\sum_x \exp\left(i \frac{2 \pi x_1}{L} \right)
         \vec{s}_x \Big|^2
\biggr\rangle
\end{equation}
is the Fourier transform of the correlation function at the lowest
non-zero momentum.
We consider several dimensionless quantities, which are also called
phenomenological couplings.
These quantities are invariant under renormalization group (RG) 
transformations. We consider the Binder cumulant $U_4$ and its sixth-order
generalization $U_6$, defined as
\begin{equation}
U_{2j} = \frac{\langle(\vec{m}^2)^j\rangle}{\langle\vec{m}^2\rangle^j} \;,
\end{equation}
where $\vec{m} = \frac{1}{V} \, \sum_x \vec{s}_x$ is the magnetization of
the system.  We also consider the ratio $R_Z = Z_a/Z_p$ of the partition
function $Z_a$ of a system with anti-periodic boundary conditions in one of the
three directions and the partition function $Z_p$ of a system with periodic
boundary conditions in all directions.  Anti-periodic boundary conditions in
$0$-direction are obtained by changing the sign of the term
$\vec{s}_x \cdot \vec{s}_y$ of the Hamiltonian for links
$\left<xy\right>$ that connect the boundaries, i.e., for $x=(L-1,x_1,x_2)$ and
$y=(0,x_1,x_2)$. In order to avoid microscopic effects at the boundary,
we require that $-\vec{s}_x$ is in the same set of values as $\vec{s}_x$.
Hence we take even values of $q$.
In the following we refer to dimensionless ratios by $R$.
Derivatives of dimensionless ratios with respect to the
inverse temperature
\begin{equation}
S_R = \frac{\partial R}{\partial K_1}
\label{sdef}
\end{equation}
are used to determine the critical exponent $\nu$. In the following these
quantities are also denoted by slope of $R$.  Note that here we decided ad 
hoc to take the partial derivative with respect to $K_1$ at fixed $K_3$ and
$D$. Taking the derivative with respect to a linear combination of $K_1$ 
and $K_3$ would also be a viable choice. For example, in Ref. \cite{myIso}, 
we pass the critical line at a fixed ratio of $K_3/K_1$.
We did not implement both the derivatives with respect to $K_1$ and $K_3$, 
to avoid the handling of additional data.

In the FSS analysis we need the observables as a function of
$K_1$ in a certain neighborhood of the critical coupling $K_{1,c}$.
To this end, we simulate at $K_{1,s}$, 
which is a good approximation of $K_{1,c}$. In order to extrapolate
in $K_1$ we compute the coefficients of the Taylor series in $K_1-K_{1,s}$
for all quantities listed above up to the third order. Note that a
reweighting analysis is not possible, since, due to the large statistics,
we performed a binning of the data already during the simulation.

\section{Corrections to scaling}
\label{Corrections_summary}
In the analysis of our data, corrections to scaling have to be taken into
account. These corrections are either caused by irrelevant scaling fields 
or are intrinsic to the observables. These intrinsic corrections 
can be related in some cases to redundant scaling fields of certain 
RG transformations. An example of such a correction is the analytic 
background in the magnetic susceptibility. The operators associated 
with the scaling fields can be classified according to the spatial
symmetry, the internal symmetry and their dimension.
For a nice discussion see Sec. 3 of Ref. \cite{He23}. 

First we discuss the corrections related to finite values
of $q$. These are well understood and the corresponding RG exponent $Y_q$ 
is rapidly decreasing with increasing $q$. We shall use values of $q$ such 
that these corrections can be safely ignored in the analysis of our data.
Note that we stay in the neighborhood of the critical point. 
Hence the dangerously irrelevant character \cite{Nelson_danger,AmPe82}
of the perturbation can be ignored.

Next we summarize results for the leading correction exponent $\omega$
and the subleading correction exponent $\omega_{NR}$.
Note that the subleading correction is  due to the breaking 
of the spatial rotational invariance by the lattice. 

\subsection{Dependence of the physics on $q$}
\label{qDependence}
Let us  summarize results on the RG eigenvalues $Y_q$ of a 
$\mathbb{Z}_q$-invariant perturbation at the $O(2)$-invariant fixed point.
For large values of $q$ one gets
\begin{equation}
\label{newdeltaq}
 \Delta_q = c_{3/2} q^{3/2}  +  c_{1/2} q^{1/2} 
 +c_0+ O(q^{-1/2}) \;,
\end{equation}
where $Y_q=3-\Delta_q$. See Ref. \cite{Alvarez} and references therein.
This has been checked numerically in Refs. \cite{Debasish,Cuomo23}.
In Ref. \cite{Cuomo23} the estimates $c_{3/2} = 0.340(1)$ and 
$c_{1/2} = 0.23(2)$ are provided, consistent with Ref. \cite{Debasish}.
The value $c_0 \simeq -0.0937$ is predicted theoretically.
Plugging in these numbers in Eq.~(\ref{newdeltaq}) we get for example
$Y_q = 2.52$, $1.81$, $0.93$, $-0.09$, $-1.22$, ..., $-5.25$, ..., $-11.83$, ..., $-19.59$ 
for $q=1$, $2$, $3$, $4$, $5$, ...,
 $8$, ...,$12$, ...,$16$. Comparing with, for example, $Y_2= 1.7639(11)$, 
$Y_3 =0.8915(20)$, $Y_4=-0.108(6)$, Ref. \cite{O234}, or $Y_2=1.76371(11)$,
Ref. \cite{che19}, we see that  
Eq.~(\ref{newdeltaq}) is reasonably accurate down to small values of $q$.
Even in the case $Y_1=y_h=2.480912(22)$, Ref. \cite{che19}, the estimate 
obtained from Eq.~(\ref{newdeltaq}) is close. 
For our purpose, it is important to note that $Y_q$ is negative for $q\ge 4$
and rapidly
decreases with increasing $q$. For example, for $q=8$, which we 
considered in Ref. \cite{myClock}, $Y_8 \approx -5.25$. In the present work,
the smallest value of $q$ we simulated at is $q=12$.

Next let us discuss whether the $(q+1)$-state clock model, for $q \ge 4$, at 
criticality, is in the basin of attraction of the $O(2)$-invariant fixed point.
It is easy to see that the 4-state clock model is equivalent to two
decoupled Ising systems by writing the two components of the field as 
$s^{(1)} = \frac{1}{2} (\sigma + \tau)$ and $s^{(2)} =
\frac{1}{2} (\sigma - \tau)$, where $\sigma,\tau \in \{-1,1\}$, see for  
example Ref. \cite{Hove03}. Hence the 4-state clock model is not 
in the basin of attraction of the $O(2)$-invariant fixed point. For
a finite value of $D$ we have no definite answer.
For $q=5$ it is demonstrated numerically in Ref. \cite{Hove03} that 
the model undergoes a second order phase transition, where the $O(2)$ symmetry
is restored. This is confirmed for example in Ref. \cite{Shao19}.
It is plausible that the same holds for $q>5$. For $q=6$ this has been 
confirmed numerically, for example, in Ref. \cite{Shao19}. For certain finite
values of $D$ this has been demonstrated for $q=6$, $7$, $8$, and $10$ 
in Ref. \cite{myClock}. 

Note that the $\mathbb{Z}_q$-invariant perturbation for $q \ge 4$ is 
dangerously irrelevant \cite{Nelson_danger,AmPe82}. 
In the low temperature phase, in the thermodynamic
limit, the magnetization might only point in one of the $q$ directions 
given by Eq.~(\ref{directions}).
However this does not affect the FSS study in the neighborhood of the 
critical point that is performed here.

For $q \ge 5$ and finite $D$ there is a range $[\infty,D_{tri})$, 
where the transition is continuous. For $D<D_{tri}$ it becomes first order.
In Ref. \cite{myClock} we find $-0.87 < D_{tri} <-0.86$ for $q=8$ and $K_3=0$. 
For larger  $q$ the value of $D_{tri}$ should virtually not change. The 
values of $D$ simulated below are clearly in the second order range.

In ref. \cite{myClock} we find, that at the level of our numerical accuracy, 
the estimate of $K_{1,c}$ becomes indistinguishable from the 
$q \rightarrow \infty$ limit for $q \gtrapprox 10$. 
In the present study we simulate $q \ge 12$.  
Hence we expect that even for non-universal quantities, the numerical 
estimates obtained here are, within the quoted errors, valid for
the $q \rightarrow \infty$ limit.

\subsection{Correction exponents for the XY universality class}
\label{XYcorrections}
In the case of the three-dimensional Ising universality class
accurate results for the dimensions of various operators obtained by using CB
are provided in Ref. \cite{SD16}. These results are corroborated by
fuzzy sphere calculations \cite{fuzzy}. For a summary of results obtained 
by using the $\epsilon$-expansion and the $1/N$-expansion for general $N$
see Ref. \cite{He23}. These results show that, at least for the leading 
corrections, the spectrum of correction exponents is sparse. 
This is actually the reason
why we can improve models by tuning a single parameter. Or as discussed
here by tuning two parameters. Field theory suggests that this qualitative
property also holds for $N>1$, where $N$ refers to the $O(N)$ symmetry of the 
theory \cite{He23}.

Let us turn to the  XY universality class.
The leading correction exponent $\omega=\Delta_{s'}-3$ is known 
from field theory, functional renormalization group, Monte Carlo simulations, 
and the CB. In \cite{myClock} we find 
$\omega=0.789(4)$, which is fully consistent with the CB estimate 
$\Delta_{s'} = 3.794(8)$  given in table 4 of Ref. \cite{che19}.
In Ref. \cite{DePo20} the authors find $\omega=0.791(8)$ by using the
functional renormalization group (FRG).
The  subleading correction is due to the violation of spatial
rotational invariance
by the simple cubic lattice. Here we are aiming at the leading correction, 
which is associated with spin $l=4$. 
From Fig. 3 of Ref. \cite{O2corrections,private}  $\omega_{NR} = 2.02548(41)$.
Note that the value of $\omega_{NR}$ is only slightly larger than that for 
the Ising universality class. Analyzing the $\epsilon$-expansion and the 
$1/N$-expansion of $\omega_{NR}$ \cite{Ma17,De98}, where the coefficients
are taken from Ref. \cite{He23}, we conclude that for the Heisenberg 
universality class $\omega_{NR}$ takes a similar value as for the 
XY universality class, and then monotonically decreases with increasing $N$, 
reaching $\omega_{NR}=2$ in the limit $N \rightarrow \infty$. Note that
$\omega_{NR}=2$  is the result for a free field theory.
Note that in the case of the Ising universality the subleading correction
exponent for $l=4$ is $\omega_{NR}'=3.42065(64)$  given in table 2 
of \cite{SD16}. In Fig. 3 of Ref. \cite{O2corrections} we see that 
$\omega_{l}$ rapidly increases with $l$. For example for $l=8$ one
gets $\omega_{l=8} \approx 6.03$.

\subsection{Dependence of the amplitudes of corrections on 
the parameters}
\label{ampDependence}
\begin{figure}
\begin{center}
\includegraphics[width=14.5cm]{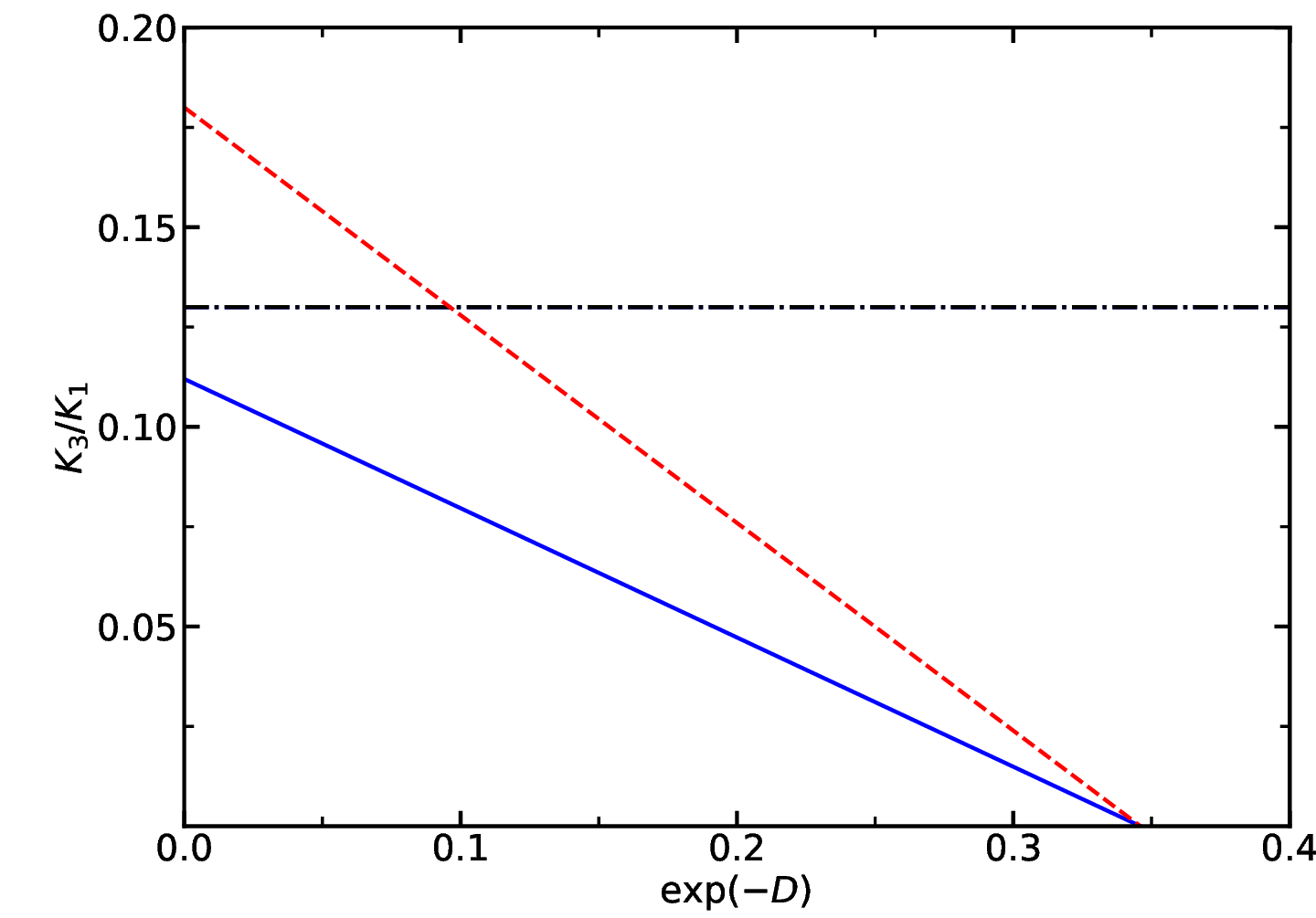}
\caption{\label{fig_sketch}
We sketch possible scenarios for the zeros of the leading correction 
amplitude $b(D,r_K)$ and the subleading correction amplitude 
$c(D,r_K)$. 
The straight dash-dotted line gives $c(D,r_K)=0$, while the straight 
solid line and the dashed line give two a priori possible behaviors 
of $b(D,r_K)=0$.
The plot resembles the qualitative behavior discussed in the text:
Since $c(D,r_K)$ depends weakly on $D$, the zero occurs at an 
approximately constant value of $r_K$. Here we have used the value
of $r_K^{iso}$ for $D \rightarrow \infty$ 
that we shall find below in Sec. \ref{Restoring}. In the case of 
$b(D,r_K)=0$ we have used $D^*\approx 1.06$ for $K_3=0$ from 
Ref. \cite{myClock} as one endpoint. 
For $D \rightarrow \infty$ we find below $r_K^*<r_K^{iso}$ 
and there is no intersection of the zeros of $b(D,r_K)$ (solid
line) and $c(D,r_K)=0$.
The dashed line gives qualitatively the result
obtained for the Blume-Capel model with 3N couplings. Here we get an 
intersection at a finite value of $D$. 
}
\end{center}
\end{figure}

Finite size scaling (FSS) determines the behavior of dimensionless quantities
at the critical point on a finite lattice. As example let us consider the 
ratio of partition functions at criticality
\begin{eqnarray}
 [Z_a/Z_p](L,D,K_{1,c},r_K) &=& [Z_a/Z_p]^* + b(D,r_K) L^{-\omega} 
     + d \; b^2(D,r_K) L^{-2 \omega} + ... \nonumber \\
&&+ c(D,r_K) L^{-\omega_{NR}} + ... \; ,  
\end{eqnarray}
where $r_K=K_3/K_1$. Here we assume a lattice with the linear lattice
size $L=L_0=L_1=L_2$ with periodic boundary conditions.
Note that RG predicts that the amplitudes of the corrections in other 
quantities are proportional to the ones of $Z_a/Z_p$, since they originate 
from the same scaling fields. Finding an
improved model means finding $(D,r_K)$ such that $b(D,r_K)=0$.
To this end, varying a single parameter is actually sufficient. 
Note that a priori it is not guaranteed that such a zero exists for the 
model studied. Here we attempt to find a point in the parameter space, 
where both amplitudes $b(D,r_K)$ and $c(D,r_K)$ vanish.
In order to develop a strategy to find these zeros, a qualitative 
understanding of the dependence of $b(D,r_K)$ and $c(D,r_K)$ on their
arguments is helpful. It is plausible that the behavior is qualitatively
the same as in the case of the Blume-Capel model with next and third next 
nearest neighbor interactions that we studied in Ref. \cite{myIso}.
The first proposition is that $c(D,r_K)$, for fixed $r_K$,  
depends very little on $D$, while $b(D,r_K)$ strongly 
depends on $D$. On the other hand, for fixed $D$, both
$b(D,r_K)$ and $c(D,r_K)$  have a clear dependence on $r_K$. 
Based on the results for the Blume-Capel model, we expect that, 
at least for small $r_K>0$, increasing $r_K$ has qualitatively
the same effect on $b(D,r_K)$ as decreasing $D$. Hence a point in the 
$(D,r_K)$ plane with $b(D,r_K)=0$ and $c(D,r_K)=0$ can 
only be found if $r_K^* \ge r_K^{iso}$ for $D \rightarrow \infty$,
where $r_K^*$ is the zero of $b$ and $r_K^{iso}$ the zero of $c$. 
In Fig. \ref{fig_sketch} we sketch the behavior of the zeros of the 
correction amplitudes discussed above.

\section{Restoring isotropy}
\label{Restoring}
In a typical FSS setting with $\xi_{\exp} \gtrapprox L$, where $\xi_{\exp}$ 
is the exponential correlation length in the thermodynamic limit, it is 
difficult to detect the violation of the rotational invariance that is caused
by the lattice.
This is due to the fact that rotational invariance is also 
broken by the torus geometry of the system with periodic boundary 
conditions. To avoid this problem, we simulate in the high temperature 
phase such that $\xi_{\exp} \ll L$. Here, the correlation function at 
distances considerably smaller than the lattice size is barely affected by
the torus geometry.

The simulations in the high temperature phase  were performed
by using the single cluster algorithm \cite{Wolff}. In the case of the 
$(q+1)$-state clock model
with finite $D$, local updates that allow the transition
from $s_x=0$ to $|s_x|= 1$ and vice versa were used in addition. For a
more detailed discussion of such a hybrid update scheme see for example
Sec. IV of Ref. \cite{myClock}. The numerical study presented below
is similar to the one discussed in Sec. IV of Ref. \cite{myIso}
for the Ising universality class in three dimensions.

Below, we first simulate $K_3=0$ at $D \approx D^*$ and
$D \rightarrow \infty$ to establish so to speak 
the baseline and to check that the behavior of the  spatial unisotropy 
is compatible with the theoretical prediction for the exponent $\omega_{NR}$.
Then, for $D \rightarrow \infty$,  we search for $r_{K}^{iso}$, where 
the leading contribution to the spatial unisotropy vanishes.

\subsection{The correlation length in the high temperature phase}
In order to quantify spatial anisotropy, we determine the correlation length
in three different directions of the lattice. Below we discuss how the
correlation length is determined. We start with the definition of the
basic quantities. 

Slice averages are given by
\begin{equation}
 \vec{S}_{(1,0,0)} (x_0) = \sum_{x_1,x_2} \vec{s}_{x_0,x_1,x_2} \;,
\end{equation}
where the slice is perpendicular to the $(1,0,0)$-axis.  In addition,
we consider slices perpendicular to the $(1,1,0)$ and the $(1,1,1)$-axis.
The corresponding slice averages are given by
\begin{equation}
\vec{S}_{(1,1,0)} (x_0)  = \sum_{x_1,x_2}   \vec{s}_{x_0-x_1,x_1,x_2}
\end{equation}
and
\begin{equation}
\vec{S}_{(1,1,1)} (x_0)   = \sum_{x_1,x_2}   \vec{s}_{x_0-x_1-x_2,x_1,x_2} \;\;.
\end{equation}
Note that the arithmetics of the coordinates is understood modulo the linear
lattice size $L$. The distance between adjacent slices is $d_s=1$, $2^{-1/2}$,
and $3^{-1/2}$ for slices perpendicular to the $(1,0,0)$-,
$(1,1,0)$- and the $(1,1,1)$-axis, respectively. Going from a site in the 
direction of the $(1,1,0)$- and the $(1,1,1)$-axis, we hit the next site 
in the second and the third next slice, respectively.
The slice correlation function is defined as
\begin{equation}
 G_r(t) = \langle  \vec{S}_r(x_0) \cdot \vec{S}_r(x_0+t) \rangle \;\;,
\end{equation}
where $r=(1,0,0)$, $(1,1,0)$, or $(1,1,1)$ and
$x_0+t$ is understood modulo the linear lattice size $L$.

In our simulations, in order to reduce the statistical error,
we average over all $x_0$ and all directions equivalent to those given by the
$(1,0,0)$-, $(1,1,0)$- and the $(1,1,1)$-axis, respectively. The correlation
function is determined by using the variance reduced estimator associated with
the cluster algorithm \cite{SwWa87,Wolff}.
We define the effective correlation length in direction $r$ by
\begin{equation}
 \xi_{r,eff}(t) =  \frac{s d_r}{\ln(G_r(t)/G_r(t+s))} \; ,
\end{equation}
where $L \gg t$ is assumed, $d_r$ is the distance between adjacent slices and
$s$ is a stride. Here we take $s=1,2,$ and $3$ for $r=(1,0,0)$, $(1,1,0)$,
and $(1,1,1)$, respectively. To relax $L \gg t$ to some extent, we take
the periodicity of the lattice into account. To this end we solve
numerically
\begin{eqnarray}
 G_r(t) &=& c \left (\exp\left(-\frac{d_r t}{\xi_{r,eff}(t)} \right)  +\exp\left(-\frac{d_r (L-t)}{\xi_{r,eff}(t)} \right)    \right) \;, \\
 G_r(t+s)&=& c \left (\exp\left(-\frac{d_r (t+s)}{\xi_{r,eff}(t)} \right)  +\exp\left(-\frac{d_r (L-t-s)}{\xi_{r,eff}(t)} \right)\right) \;.
\end{eqnarray}
Similar to the Ising universality class in three dimensions,
in the high temperature phase $\xi_{r,eff}(t)$ converges quickly as
$t \rightarrow \infty$. See Ref. \cite{XYamp,MyVar} and references therein.

In a set of preliminary simulations, we determined the lattice size $L$ and
distance $t$ that is needed to keep deviations from the desired limit
$L \rightarrow \infty$ followed by $t \rightarrow \infty$ at a size smaller
than the statistical error.
We conclude that $d_r t \simeq 2 \xi_r$ and $L \simeq 20 \xi_r$ is sufficient.
In the following we take $\xi_{r,eff}(t)$ at $d_r t \simeq 2 \xi_r$ as estimate
of the correlation length $\xi_r$.  
In order to quantify the spatial anisotropy, we study the ratios
\begin{equation}
\label{xiratios}
r_2=\frac{\xi_{(1,0,0)}}{\xi_{(1,1,0)}} \;\;,\;
r_3=\frac{\xi_{(1,0,0)}}{\xi_{(1,1,1)}}
\end{equation}
in the neighborhood of the critical point. 
Note that for the ratios
$r_2$ and $r_3$, the covariance between the correlation lengths
in the different directions are properly taken into account by performing
a Jackknife analysis.

We expect
\begin{equation}
r_i-1 = a_i(D,r_K)  \xi_{(1,0,0)}^{-\omega_{NR}} + b_i(D,r_K)  \xi_{(1,0,0)}^{-\omega_{NR}'} + ... \;.
\end{equation}

\subsection{Numerical results for the $(q+1)$-state clock model at $K_3=0$}
First we determine the correlation length in three directions for the 
$(q+1)$-state clock model at $K_3=0$. 
In our simulations we take $q=16$. We simulated at
$D=1.05$ and $1.07$, which are close to $D^*$. In Ref. \cite{myClock} we find
 $D^* =  1.058(13)$ for $q=8$. For larger values of $q$, $D^*$ should change 
only by little. 
In particular $|D^*(q=10) - D^*(q=8)| \le 0.0005$ is estimated in Ref.
\cite{myClock}. 
Our numerical results for the correlation length are summarized in 
table \ref{qp1}.  
\begin{table}
\caption{\sl \label{qp1}
We give estimates of the correlation length $\xi_{(1,0,0)}$
and the ratios $r_2$ and $r_3$ for the $(16+1)$-state clock model with 
nearest neighbor couplings only at $D=1.05$, $1.07$ and $\infty$. 
$L$ is the linear lattice size and $K_1$ the nearest neighbor coupling.
For a discussion see the text.
}
\begin{center}
\begin{tabular}{llrlllll}
\hline
\mc{1}{c}{$D$} &
\mc{1}{c}{$K_1$} &  \mc{1}{c}{$L$}  &
\mc{1}{c}{$\xi_{(1,0,0)}$}  &
\mc{1}{c}{$r_2$} & \mc{1}{c}{$r_3$} & 
\mc{1}{c}{$(r_2-1) \xi_{(1,0,0)}^{2.02548} $} &
\mc{1}{c}{$(r_3-1) \xi_{(1,0,0)}^{2.02548} $} \\
\hline
1.05&0.53185 &60 &3.01417(5)&1.002258(6)&1.003022(7)&0.02110(6)& 0.02824(7) \\
1.05&0.54186 &80 &4.02463(7)&1.001263(6)&1.001698(8)&0.02120(10)&0.02850(12)\\
1.05&0.547185&100&5.03384(9)&1.000800(5)&1.001076(6)&0.02112(13)&0.02841(16)\\
1.05&0.55042 &120&6.04634(7)&1.000562(3)&1.000759(4)&0.02151(12)&0.02905(15)\\
\hline
1.07&0.5398 &80&  4.00873(5)&1.001272(5)&1.001705(6)& 0.02118(8) &0.02839(10)\\
\hline
$\infty$&0.4258&60&3.01208(4)&1.002254(5)&1.003016(6) &0.02103(5) &0.02814(6)\\
$\infty$&0.4353&80&4.00646(5)&1.001276(6)&1.001713(7) &0.02122(10)&0.02849(12)\\
$\infty$&0.44049&100&5.00670(5)&1.000810(5)&1.001090(6)&0.02115(13)&0.02847(16)\\
\hline
\end{tabular}
\end{center}
\end{table}
For example for $D \rightarrow \infty$, $K_1=0.44049$ we performed about 
$2 \times 10^{11}$ single cluster updates.

As expected, for $D=1.05$, we see that $(r_i-1) \xi_{(1,0,0)}^{2.02548}$ stays 
roughly constant with increasing correlation length.
The estimates of $r_i$ for $\xi_{(1,0,0)} \approx 4$ for $D=1.05$ and $1.07$
are consistent. The estimates of $(r_i-1) \xi_{(1,0,0)}^{2.02548}$ for 
$D \rightarrow \infty$ and $D=1.05$ are equal within the statistical error. 
This confirms the hypothesis that the violation of isotropy depends little
on $D$.

\subsection{Numerical results for the $q$-state clock model at $K_3>0$}
Next we studied the 3N $q$-state clock model. As above, we take $q=16$.
Here we aim at finding $r_K^{iso}=K_3/K_1$, where rotational invariance is 
restored.

Based on preliminary simulations, we have chosen $K_1$ and $K_3$ such that
$\xi_{(1,0,0)} \approx 3$, $4$, $5$, or $6$. Furthermore  $r_K$ is chosen 
such that  the deviation of $r_{2}$ and $r_{3}$ from one is small.  
Our numerical results are summarized in table \ref{qp2}. Throughout
the linear lattice size is chosen such that $L \approx 20 \xi_{(1,0,0)}$
except for $(K_1,K_3)  =(0.3372,0.044)$, where we have simulated $L=80$
in addition to $L=60$ to check that finite size effects are not larger 
than the statistical error.

In the first step of the analysis, we determine the zeros of $r_2-1$ and
$r_3-1$ for each value of $\xi_{(1,0,0)}$. To this end, we performed a
linear fit of $r_i-1$ in $r_K$. In the case of $r_2-1$, we get
$r^{iso}_K =0.12744(29)$, $0.12919(36)$, $0.12832(60)$, and $0.12926(87)$ 
for $\xi_{(1,0,0)} \approx 3, 4, 5$, and $6$, respectively.
The corresponding numbers for $r_3-1$ are
$0.12729(25)$, $0.12950(32)$, $0.12953(52)$, and $0.13107(74)$. 
We note that
the estimates obtained from $r_2$ and $r_3$ are consistent. We expect that
corrections decay as $\xi_{(1,0,0)}^{-\omega_{NR}'+\omega_{NR}}$. 
For the critical limit, we arrive at
\begin{equation}
\label{RK}
r_K^{iso}=0.132(2) \;,
\end{equation} 
where the error is chosen such that it covers both the estimates obtained
from $r_2$ and $r_3$.
Note that for the free field theory one gets $r_K^{iso}=0.125$ \cite{PaKa05} 
and $r_K^{iso}=0.129(1)$ for the Blume-Capel model with nearest neighbor
and next-to-next-to-nearest neighbor interactions \cite{myIso}.

\begin{table}
\caption{\sl \label{qp2}
Estimates of the correlation length $\xi_{(1,0,0)}$
and the ratios $r_2$ and $r_3$ for the $16$-state clock model with next and 
next-to-next-to-nearest neighbor coupling constants.
$L$ is the linear lattice size and $K_1$ and $K_3$ are the
coupling constants. For a discussion see the text.
}
\begin{center}
\begin{tabular}{lllrlll}
\hline
\mc{1}{c}{$K_1$} &
\mc{1}{c}{$K_3$} &  
\mc{1}{c}{$K_3/K_1$} &  
\mc{1}{c}{$L$}  &
\mc{1}{c}{$\xi_{(1,0,0)}$}  &
 \mc{1}{c}{$r_2$} & \mc{1}{c}{$r_3$} \\
\hline                
0.347  & 0.039 &0.1123919... & 60 & 3.02338(3) & 1.000187(5) & 1.000247(5) \\
0.3448 & 0.04  &0.1160092... & 60 & 3.00471(3) & 1.000140(5) & 1.000186(6) \\
0.341  &0.042  &0.1231671... & 60 & 3.01011(3) & 1.000054(5) & 1.000067(6) \\
0.3372 & 0.044 &0.1304863... & 60 & 3.01430(3) & 0.999962(5) & 0.999948(6) \\
\hline
0.3372 & 0.044 & 0.1304863... & 80 & 3.01431(2) & 0.999962(4)& 0.999944(5) \\
\hline
0.3548 & 0.04  & 0.1127395... &80 & 4.02702(3) & 1.000100(5)& 1.000145(6) \\
0.347  & 0.044 & 0.1268011... &80 & 4.01415(3) & 1.000013(4) & 1.000021(5) \\
0.34556& 0.0447& 0.1293552... &80 & 4.00013(3) & 1.000001(4) & 1.000003(5) \\
0.34316& 0.046 & 0.1340482... &80 & 4.01300(3) & 0.999970(4) & 0.999961(5) \\
\hline
0.36   & 0.04  & 0.1111111... &100 & 5.02203(4) & 1.000080(4) & 1.000112(5)  \\
0.3582 & 0.041 & 0.1144611... &100 & 5.05559(4) & 1.000062(4) & 1.000088(5)  \\
0.3523 & 0.044 & 0.1248935... &100 & 5.02358(4) & 1.000013(4) & 1.000026(5) \\
0.3484 & 0.046 & 0.1320321... &100 & 5.00579(3) & 0.999985(4) & 0.999986(4) \\
\hline
0.3516 & 0.046  & 0.1308304... & 120 & 5.99989(3) & 0.999995(3) & 1.000001(3) \\
0.3621 &0.04054 & 0.1119580... & 120 & 6.00411(4) & 1.000055(4) & 1.000079(4) \\
\hline
\end{tabular}
\end{center}
\end{table}

\section{Simulations at criticality}
\label{FSSnumerics}
In this section we discuss the simulations performed close to the critical
point, such that $\xi \gg L$.  We analyze data for $q=8$, $K_3=0$ at $D=1.02$,
$1.05$, and $1.07$. These were already generated for
Ref. \cite{myClock}. For $q=12$, $K_3=0$ we generated new data at $D=1.05$, 
$1.06$, and $1.07$. 
Our main focus is on $D \rightarrow \infty$, $K_3>0$. Here $q\ge 24$ 
throughout. We simulated at $K_3=0.04$, $0.0415$, $0.042$,
$0.046$, and $0.05$. These values of $D$ and $K_3$ are close to $D^*$ and
$K_3^*$, respectively. In the following, we briefly discuss the algorithm that
has been used for the simulations. Then we summarize the lattice
sizes that have been simulated and the statistics that we achieved
in our simulations.

First we analyze the dimensionless quantities $Z_a/Z_p$,
$\xi_{2nd}/L$, $U_4$ and $U_6$, defined in Sec. \ref{def}. We obtain
estimates of the fixed point values of these  dimensionless quantities
and estimates of the zeros of the leading correction amplitude
$D^*$ and $K_3^*$. We confirm the estimate of $r_K^{iso}$ obtained in the 
previous section. We get accurate estimates of $K_{1,c}(D,K_3)$.

Then we compute the critical exponent $\eta$ by analyzing the magnetic 
susceptibility $\chi$ at criticality and the RG exponent $y_t=1/\nu$
by analyzing slopes $S_R=\frac{\partial R}{\partial K}$ 
of dimensionless quantities at criticality.

\subsection{The simulation program/algorithm}

We simulated the model by using a hybrid of the Metropolis, the single cluster
\cite{Wolff} and the wall cluster algorithm \cite{HaPiVi}. For
finite $D$, the Metropolis updates are needed to update $\vec{s}_x=(0,0)$ to
$|\vec{s}_x|=1$ and vice versa. For a discussion see Sec. IV of 
Ref. \cite{myClock}. In the case of $K_3>0$ and $D \rightarrow \infty$, 
we kept the Metropolis updates, even though they are not required.
During the study the precise composition of the update cycle changed. 
The advantage of simulating with finite $q$ is that delete probabilities 
for the cluster algorithm and acceptance probabilities for the Metropolis
algorithm can be computed at the beginning of the simulation and be stored
in a lookup table. Furthermore, for the values of $q$ considered here, 8 bit 
\verb+char+ variables are sufficient to store the spins. Note
that the efficiency of the hybrid update does not depend sharply on the
precise composition of the update cycle. In our most recent simulations 
for $K_3>0$ and $D \rightarrow \infty$ we have used
\begin{verbatim}
for(i=0;i<N_bin;i++)
  {
  for(k=0;k<3;k++)
    {
    metropolis();
    for(l=0;l <L/4;l++) single_cluster();
    2 times wall_cluster(direction=k);
    measurements();
    }
  }
\end{verbatim}

In the pseudo-C code given above, \verb+metropolis()+ is a sweep
with the local Metropolis over the lattice. 
\verb+single_cluster()+ is one single cluster update. The average size
of a single cluster, at the critical point, grows like $L^{2-\eta}$. 
Hence performing $L/4$ single cluster updates in one update cycle, 
the fraction of updated sites stays roughly constant with increasing 
linear lattice size $L$. \verb+wall_cluster(direction=k)+ is the wall 
cluster update. It is performed for planes perpendicular to the 
$(1,0,0)$, $(0,1,0)$, and $(0,0,1)$-axis in a sequence.
In order to compute $Z_a/Z_p$ we need two
subsequent wall cluster updates, where the two reflection axes are
perpendicular. In Ref. \cite{XY1} the authors find that for the 
ddXY model the average size $W$ of the wall cluster behaves as $W=a L^{2.512}$,
where the error of the exponent is not given. 
Here, for $K_3=0.0415$ and $K_1 \approx K_{1,c}$ we get 
$W=a L^{2.4828(3)} + b  L^{1.863(6)}$
with acceptable fits starting from $L \approx 20$. 
We made no effort to estimate systematic 
errors in the values of the exponents. 
This result means that the fraction of time spent with the 
wall cluster update decreases with increasing lattice size.

In the major part of the simulations 
we have used a hybrid of generators, where
one component is the \verb$ xoshiro256+ $ taken from \cite{VignaWWW}.
For a discussion of the generator see \cite{ViBl18}.
As second component we used
a 96 bit linear congruential generator with the multiplier and the
increment $a=c=$\verb+0xc580cadd754f7336d2eaa27d+ and the modulus $m=2^{96}$
suggested by O'Neill \cite{ONeill_minimal}. In this case we used our own
implementation. The third component is a multiply-with-carry generator taken 
from \cite{KISS_wiki}.
For a more detailed discussion see the Appendix A of Ref. \cite{myIso}.

In preliminary runs with about $10^6$ update cycles at our preliminary
estimate of $K_{1,c}$ for $K_{3}=0.0415$ we determined the integrated 
autocorrelation times 
of the energy density and the magnetic susceptibility. For the update cycle 
given above, in units of measurements, we find for example that
$\tau_{int,ene} = 3.45(3)$ and $6.16(8)$ for $L=40$ and $200$, respectively.
The integrated autocorrelation time of the magnetic susceptibility is 
smaller:
$\tau_{int,\chi}=2.94(3)$ and $3.73(5)$ for $L=40$ and $200$, respectively.
The autocorrelation times grow slowly with increasing lattice size. 
Since the fraction of sites that are updated by the 
wall cluster shrinks with increasing linear lattice size, we abstain from 
quoting an estimate of the dynamical critical exponent $z$.

\subsection{The data}
For $D \rightarrow \infty$, we have generated  data for 
$K_3=0.04$, $0.0415$, $0.042$, $0.046$, and $0.05$.
The simulations were performed at iteratively improved estimates of 
$K_{1,c}(K_3)$. The highest statistics is achieved for $K_3=0.0415$, 
which is close to our final estimate of $K_3^*$.  Here we simulated the 
linear lattice sizes $L=8$, $9$, ..., $20$, $22$, ..., $36$, $40$, ..., 
$72$, $80$, $90$, $100$, $120$, $140$, $160$, and $200$. 
For example for $L=24$, $64$, $100$, 
$140$, and $200$, we performed about $1.4 \times 10^{10}$, $3.3  \times 10^9$,
$8.9 \times 10^8$, $6.4 \times 10^8$, and $3 \times 10^8$ measurements.
The statistics is monotonically decreasing with increasing lattice size, 
while the CPU time that is used is increasing with the lattice size.
In total we used roughly the equivalent of 60 years on a single
core of an Intel(R) Xeon(R) CPU E3-1225 v3 running at 3.20 GHz
for the simulations at $K_3=0.0415$.  

By mistake, for some of the runs at $K_3=0.042$ and $0.046$ we did not 
compute $\xi_{2nd}/L$ correctly. This actually triggered us to 
perform a joint analysis of $Z_a/Z_p$ and $U_4$ only, instead of
$Z_a/Z_p$, $\xi_{2nd}/L$, $U_4$ and $U_6$ in the section below.

For $q=12$ and $K_3=0$, we simulated at $D=1.05$, $1.06$, and $1.07$.
First simulations, already performed a few years back, were performed by 
using the SIMD-oriented Fast Mersenne Twister (SMFT) \cite{twister}
pseudo-random number generator, where SIMD is the abbreviation for
single instruction, multiple data.
For $D=1.05$ and $1.07$ we performed simulations for the linear lattice
sizes $L=5$, $6$, ..., $14$, $16$, ..., $24$, $28$, ..., $40$, $48$, ..., $72$.
For $D=1.06$ we simulated in addition $L=15$, $26$, $30$, $44$, $80$, and 
$100$.
For $D=1.06$,  up to $L=28$,
we performed about $7 \times 10^{9}$ measurements. The number of measurement
is then decreasing to $6 \times 10^{8}$ for $L=100$.
In total the equivalent of about 20 years on a single core of an Intel(R) 
Xeon(R) CPU E3-1225 v3 running at 3.20 GHz are used for these simulations.

\subsection{Finite size scaling of dimensionless quantities}
\label{dimless}
We analyze dimensionless quantities for $q=8$ at $D=1.02$, $1.05$ and $1.07$, 
and $q=12$ at $D=1.05$, $1.06$ and $1.07$,
where in all cases $K_3=0$, and  $K_3=0.04$, $0.0415$, $0.042$, $0.046$,
and $0.05$  for $D \rightarrow \infty$. The data for $q=8$, $K_3=0$
were generated for Ref. \cite{myClock}. 
For a discussion of the theoretical background of the FSS analysis
performed here see Refs. \cite{XY1,XY2,myClock}. First we jointly fit the
ratio of partition functions $Z_a/Z_p$ and the Binder cumulant $U_4$.
We use Ans\"atze of the form
\begin{equation}
\label{masterR}
\left . R_{i}(L,D,K_1,K_3) \right |_{K_1=K_{1,c}} 
   = R_i^* + \sum_{j=1}^{j_{max}} c_{i,j}(D,K_3) L^{-\epsilon_j} \;.
\end{equation}
Note that we have simulated at
$K_{1,s} \approx K_{1,c}$. In addition to the value of
$R_i(L,D,K_{1,s},K_3)$, we determine the Taylor coefficients of the
expansion of $R_i$ in $(K_1-K_{1,s})$ up to the third order. In our
fits, we keep $R_i(L,D,K_{1,s},K_3)$ on the right side of the equation,
and bring the terms proportional to $(K_1-K_{1,s})^{\alpha}$ for
$\alpha=1$, $2$, and $3$ to the left. Furthermore, we ignore the statistical
error of the Taylor coefficients. This way, we can treat $K_{1,c}$ as a
free parameter in the fit.

We perform fits with $j_{max}=3$, where we take
the correction exponents $\epsilon_1=\omega$, 
$\epsilon_2=2-\eta$, and $\epsilon_3=\omega_{NR}$,  with the
numerical values $\omega=0.789$,  
$\eta=0.03817$, and $\omega_{NR}=2.02548$.
For a discussion 
see Sec. \ref{XYcorrections}. As check we performed fits with $j_{max}=4$,
taking $\epsilon_4=3.6$.
We ignore corrections  proportional to $L^{-2 \omega}$ or 
$L^{-\omega-\omega_{NR}}$, since for the values of $D$ and $K_3$
that we simulate at, $|b_i(D,K_3)|$ is small.

Let us discuss the parameterization of $c_{i,j}(D,K_3)$. 
First we note that $Z_a/Z_p$ is not affected by the analytic background 
of the magnetization. Hence $c_{Z_a/Z_p,2}=0$. Furthermore, RG predicts that
$c_{i,j}(D,K_3)=r_{ilj} c_{l,j}(D,K_3)$, since they originate from 
the same scaling field.

RG predicts that the amplitudes $c_{i,j}(D,K_3)$ are smooth functions.
Here we consider values of $D$ and $K_3$ in a small range around 
$D^*$ and $K_3^*$, respectively. Therefore we approximate $c_{i,j}(D,K_3)$ 
by low order Taylor approximations, for the two cases $K_3=0$ and 
$D \rightarrow \infty$. Our first parameterization for $j_{max}=3$ is \\
for $K_3=0$: 
\begin{eqnarray}
c_{U_4,1}(D,K_{1,c},0)  &=& a_1 (D-D^*) \;, \;\;  
c_{Z_a/Z_p,1} = r_{Z_a/Z_p,U_4,1} \; c_{U_4,1} \nonumber\\
c_{U_4,2}(D,K_{1,c},0)  &=& b_1 \nonumber\\
c_{Z_a/Z_p,3}(D,K_{1,c},0)  &=& d_1 \;, \;\; c_{U_4,3} =  r_{U_4, Z_a/Z_p,3} \; c_{Z_a/Z_p,3}
\end{eqnarray}
and for $D \rightarrow \infty$: \\
\begin{eqnarray}
c_{U_4,1}(\infty,K_{1,c},K_3)  &=& a_2 (K_3-K_3^*) + e_2 (K_3-K_3^*)^2\;, \;\;   c_{Z_a/Z_p,1} = r_{Z_a/Z_p,U_4,1} \; c_{U_4,1}
\label{para4} \nonumber \\
c_{U_4,2}(\infty,K_{1,c},K_3)  &=& b_2  \nonumber \\
c_{Z_a/Z_p,3}(\infty,K_{1,c},K_3)  &=& d_2 (K_3-K_3^{iso}) \;, \;\; c_{U_4,3} =  r_{U_4, Z_a/Z_p,3} \; c_{Z_a/Z_p,3} 
\end{eqnarray}
In the fits, the free parameters are $K_{1,c}(D,K_3)$, $R_i^*$, $D^*$, 
$K_3^*$, $K_3^{iso}$, $a_1$, $a_2$,
$b_1$, $b_2$, $d_1$, $d_2$, $e_2$, $r_{Z_a/Z_p,U_4,1}$, and 
$r_{U_4, Z_a/Z_p,3}$. Fits with this parameterization are labeled by A.
Note that based on the results of Ref. \cite{myClock}, the difference 
between $K_{1,c}(D,K_3=0)$ for $q=8$ and $12$, is small but clearly larger
than the statistical error that we achieve here. Therefore, we use a 
free fit parameter for each value of $q$. For all other parameters of the 
fit, we expect that the difference is much  smaller than the statistical
error. Therefore joint fit parameters are used.

In a second parameterization, we have added a linear dependence of 
$c_{U_4,2}$ on $D$ and
$K_3$, respectively. Fits with this parameterization are labeled by B.

For $j_{max}=4$, we take four free parameters for
the amplitudes $c_{Z_a/Z_p,4}$ and $c_{U_4,4}$ for $K_3=0$ and 
$D \rightarrow \infty$ in addition.  Putting this on top of the 
parameterization A is denoted by AA below, while putting it on top of
B is denoted by AB.

We fitted the data taking into account linear lattice sizes $L \ge L_{min}$.
Typically, first $\chi/$D.O.F. decreases rapidly, with increasing $L_{min}$
and then levels off at $\chi/$D.O.F. $\approx 1$. The corresponding $p$-values 
reach an acceptable range. For $L_{min} \approx 13$, $13$, $9$, and $9$ the 
levelling off is reached for the parameterizations A, B, AA, and AB, 
respectively. In the case of AA and BB, $\epsilon_4=3.6$ is used.
For a discussion of  $\chi/$D.O.F. and the corresponding $p$-value see 
appendix A of Ref. \cite{myCubic2}.
Throughout this work, least square fits were performed by using the function
\verb+curve_fit()+
contained in the SciPy  library \cite{pythonSciPy}.
Plots were generated by using the Matplotlib library \cite{plotting}.

\begin{figure}
\begin{center}
\includegraphics[width=14.5cm]{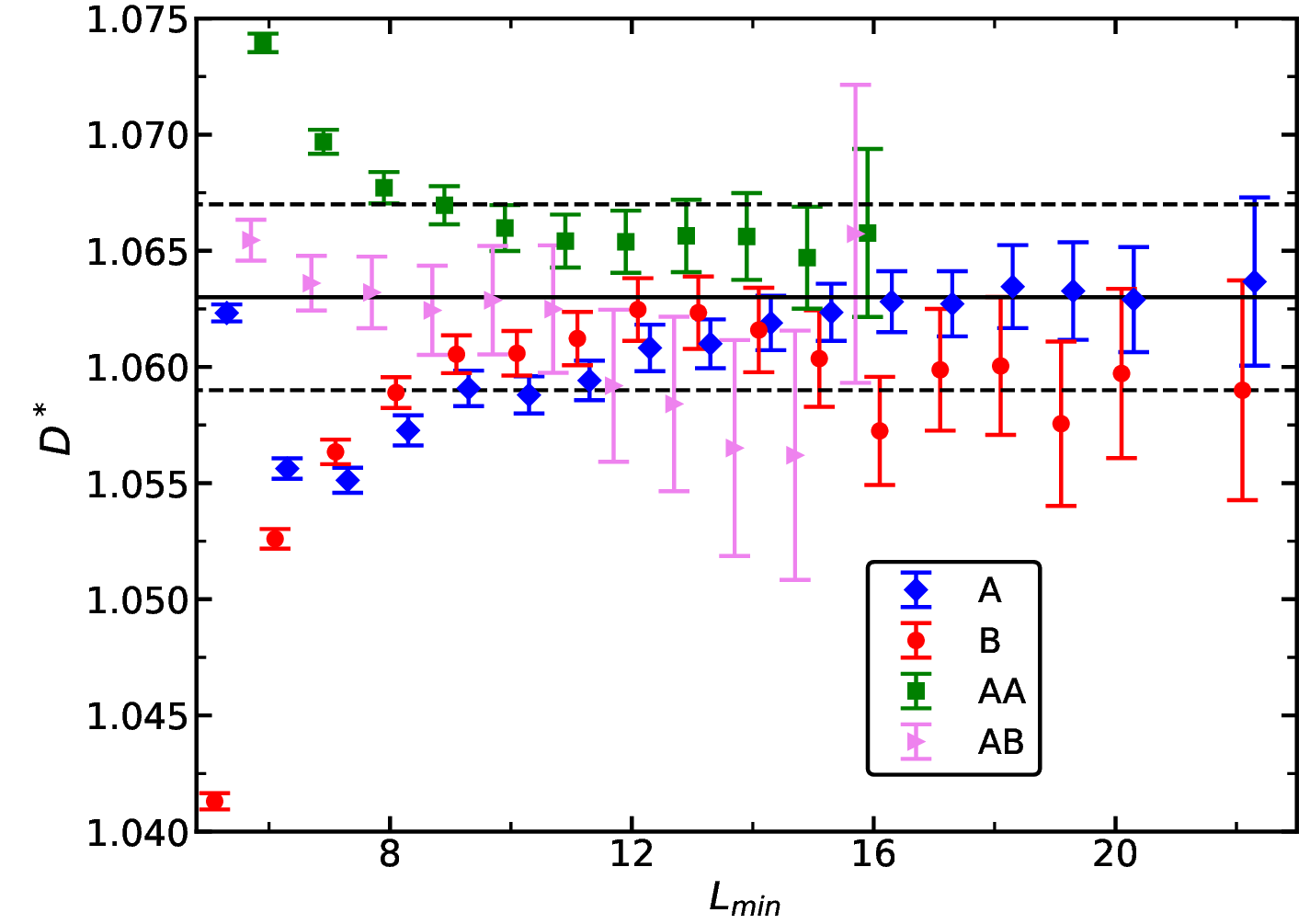}
\caption{\label{Dstar}
We plot estimates of $D^*$ for $K_3=0$ obtained by using four
different Ans\"atze as function of $L_{min}$, where all data obtained
for $L \ge L_{min}$ are taken into account in the fit. The Ans\"atze
$A$, $B$, $AA$ and $AB$ are discussed in the text. 
The solid line gives our final estimate $D^*=1.063$, while the dashed lines 
indicate our error estimate.
Note that the values on the $x$-axis are slightly shifted to reduce overlap
of the symbols.
}
\end{center}
\end{figure}

\begin{figure}
\begin{center}
\includegraphics[width=14.5cm]{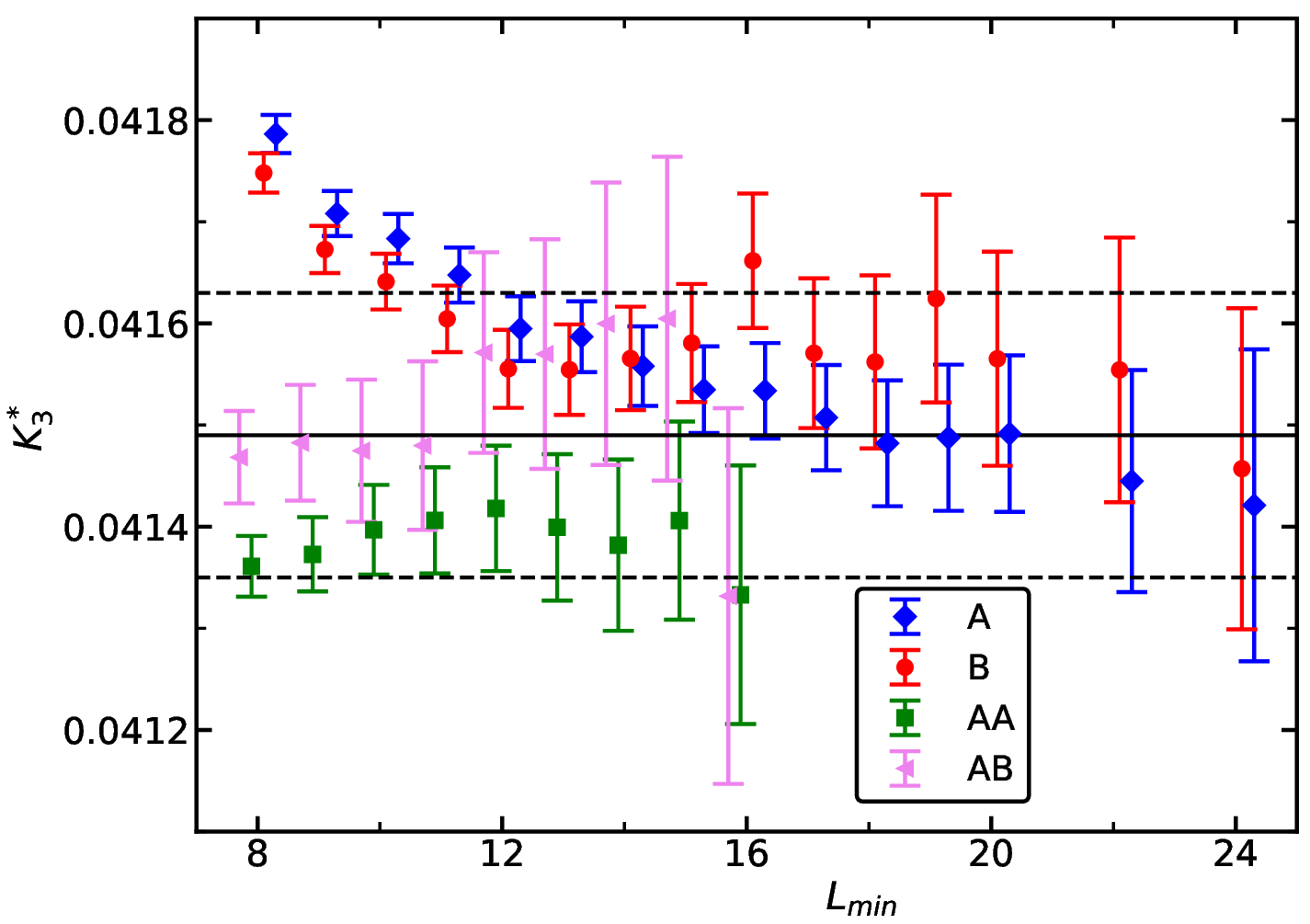}
\caption{\label{Kstar}
We plot estimates of $K_3^*$ for $D \rightarrow \infty$ obtained by 
using four 
different Ans\"atze as a function of $L_{min}$, where all data obtained
for $L \ge L_{min}$ are taken into account in the fit. The Ans\"atze labelled by
$A$, $B$, $AA$, and $AB$ are discussed in the text. 
The solid line gives our final estimate, while the dashed lines indicate our
preliminary error estimate.
Note that the values on the $x$-axis are slightly shifted to reduce overlap
of the symbols.
}
\end{center}
\end{figure}

\begin{figure}
\begin{center}
\includegraphics[width=14.5cm]{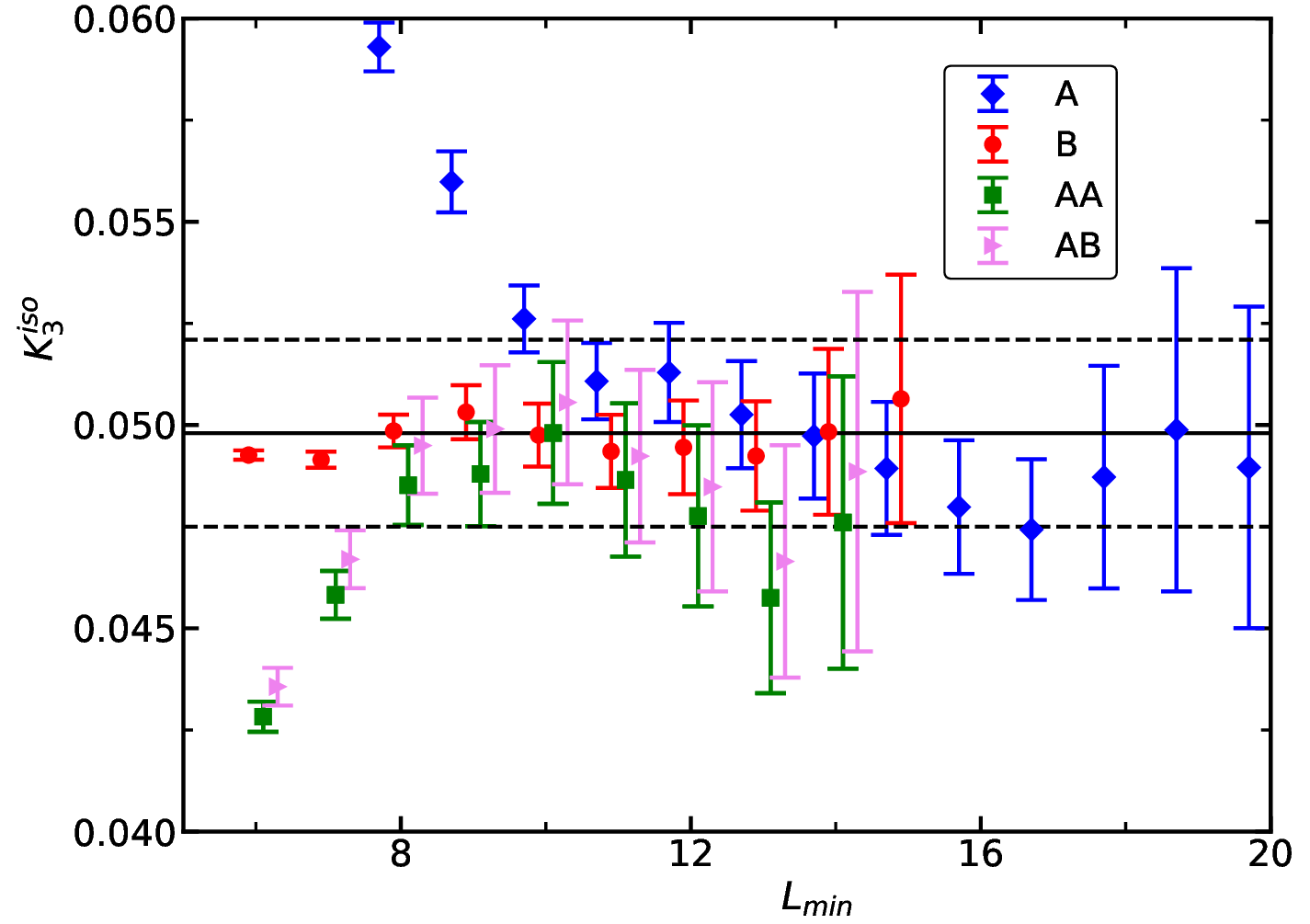}
\caption{\label{Kiso}
We plot estimates of $K_3^{iso}$ for $D \rightarrow \infty$ obtained by
using four
different Ans\"atze as a function of $L_{min}$, where all data obtained
for $L \ge L_{min}$ are taken into account in the fit. The Ans\"atze labelled by
$A$, $B$, $AA$, and $AB$ are discussed in the text.
The solid line gives our final estimate, while the dashed lines indicate our
preliminary error estimate.
Note that the values on the $x$-axis are slightly shifted to reduce overlap
of the symbols.
}
\end{center}
\end{figure}

Our final estimates of $R_i^*$, $D^*$, $K_3^*$, 
$K_3^{iso}$, and $K_{1,c}(D,K_3)$, are obtained in the following way:
We require that for each of the Ans\"atze, there is at least one $L_{min}$ with
an acceptable $p$-value that gives an estimate consistent with the final estimate.
In general, this way the error is larger than the error based on a single Ansatz.
The variation of the estimate over different Ans\"atze provides an estimate of 
systematic errors. 
In Fig. \ref{Dstar} we plot estimates of $D^*$ for $K_3=0$ as a function of 
$L_{min}$ for the four Ans\"atze discussed above. In Figs. \ref{Kstar} and
\ref{Kiso} we plot
analogous data for $K_3^*$ and $K_3^{iso}$ at $D \rightarrow \infty$, respectively. 

Our final estimates are 
\begin{eqnarray}
(Z_a/Z_p)^* &=& 0.320380(8) \\
U_4^* &=& 1.242934(10) \\
\label{ratioU}
r_{Z_a/Z_p,U_4,1}&=&-0.415(15) \\
D^* &=& 1.063(6) \\
K_3^* &=& 0.04149(14) \\
K_3^{iso} &=& 0.0498(23)  \;.
\end{eqnarray}
In terms of the ratio $r_K=K_3/K_1$ at criticality we get $r_K^*=0.1119(5)$  
and $r_K^{iso} = 0.140(8)$. The estimate of $r_K^{iso}$ is consistent
with Eq.~(\ref{RK}) obtained above but less precise. 
Furthermore one should note that 
here $K_3^{iso}$ highly relies on the fact that $Z_a/Z_p$, 
in contrast to $U_4$, $U_6$ and $\xi_{2nd}/L$, is not
affected by the analytic background of the magnetic susceptibility.
Analyzing the numbers of table \ref{qp1} and \ref{qp2}, we find that
the violation of the rotational invariance at $r_K^*$ is reduced 
by a factor of about 9 compared with $r_K=0$, which translates to 
a scale factor of a little bit less than 3. In terms of reduction of 
the volume this means a factor of about $3^3=27$. 

\begin{table}
\caption{\sl \label{betac}
Estimates of the critical coupling $K_{1,c}$ for given $D$ and $K_3$.   
}
\begin{center}
\begin{tabular}{lll}
\hline
\mc{1}{c}{$D$} &
\mc{1}{c}{$K_3$} &
\mc{1}{c}{$K_{1,c}$} \\
\hline
  1.02, $q=8$ &  0   & 0.563796225(35)  \\
  1.05, $q=8$ &   0  & 0.560823902(25)  \\
  1.05, $q=12$ &   0 & 0.560824190(30) \\
  1.06, $q=12$ &   0 & 0.559849876(22)  \\
  1.07, $q=8$ &   0  & 0.558883385(25)  \\
  1.07, $q=12$ &   0 & 0.558883648(30)  \\
\hline
$\infty$  & 0.04   & 0.37360041(3) \\
$\infty$  & 0.0415 & 0.37069947(2) \\
$\infty$  & 0.042  & 0.36973446(3) \\
$\infty$  & 0.046  & 0.36204960(4) \\
$\infty$  & 0.05   & 0.35442761(4) \\
\hline
\end{tabular}
\end{center}
\end{table}

Analyzing the data sets with $\xi_{2nd}/L$ correctly measured gives us
$\xi_{2nd}/L=0.592363(12)$ and $U_6=1.75035(5)$.
Furthermore $r_{\xi_{2nd}/L,U_4,1}  = 0.435(15)$ and $r_{U_6,U_4,1} =3.605(5)$.
Estimates of the critical coupling $K_{1,c}$ are summarized in table 
\ref{betac}. 
The ratio of $K_{1,c}$, which is denoted by $\beta_c$ in Ref.
 \cite{myClock}, for $q=8$ and $q=12$ at $D=1.07$ in table 
VII of Ref. \cite{myClock} is fully consistent with the one obtained here.
Our results for the fixed point values of the dimensionless
quantities are consistent with those given in Ref. \cite{myClock}. The accuracy
is clearly improved.

\subsection{The critical exponent $\eta$}
At the critical point, the magnetic susceptibility behaves as
\begin{eqnarray}
\chi &=& a(D,K_3) L^{2-\eta} [1 + c(D,K_3) L^{-\omega} 
      +d c^2(D,K_3) L^{-2 \omega} + ...
+
  e(D,K_3) L^{-\omega_{NR}} + ...] \nonumber \\ && + b(D,K_3) \; ,
\end{eqnarray}
where $b(D,K_3)$ is the analytic background of the magnetic susceptibility.
For a derivation of this behavior see for example Sec. III B of Ref. 
\cite{myClock}.
Instead of computing the magnetic susceptibility at an estimate of 
$K_{1,c}$, we compute it at a fixed value of a dimensionless 
quantity. In particular, here we take it at $Z_a/Z_p=0.32038$ or 
$\xi_{2nd}/L=0.592363$. 

\begin{figure}
\begin{center}
\includegraphics[width=14.5cm]{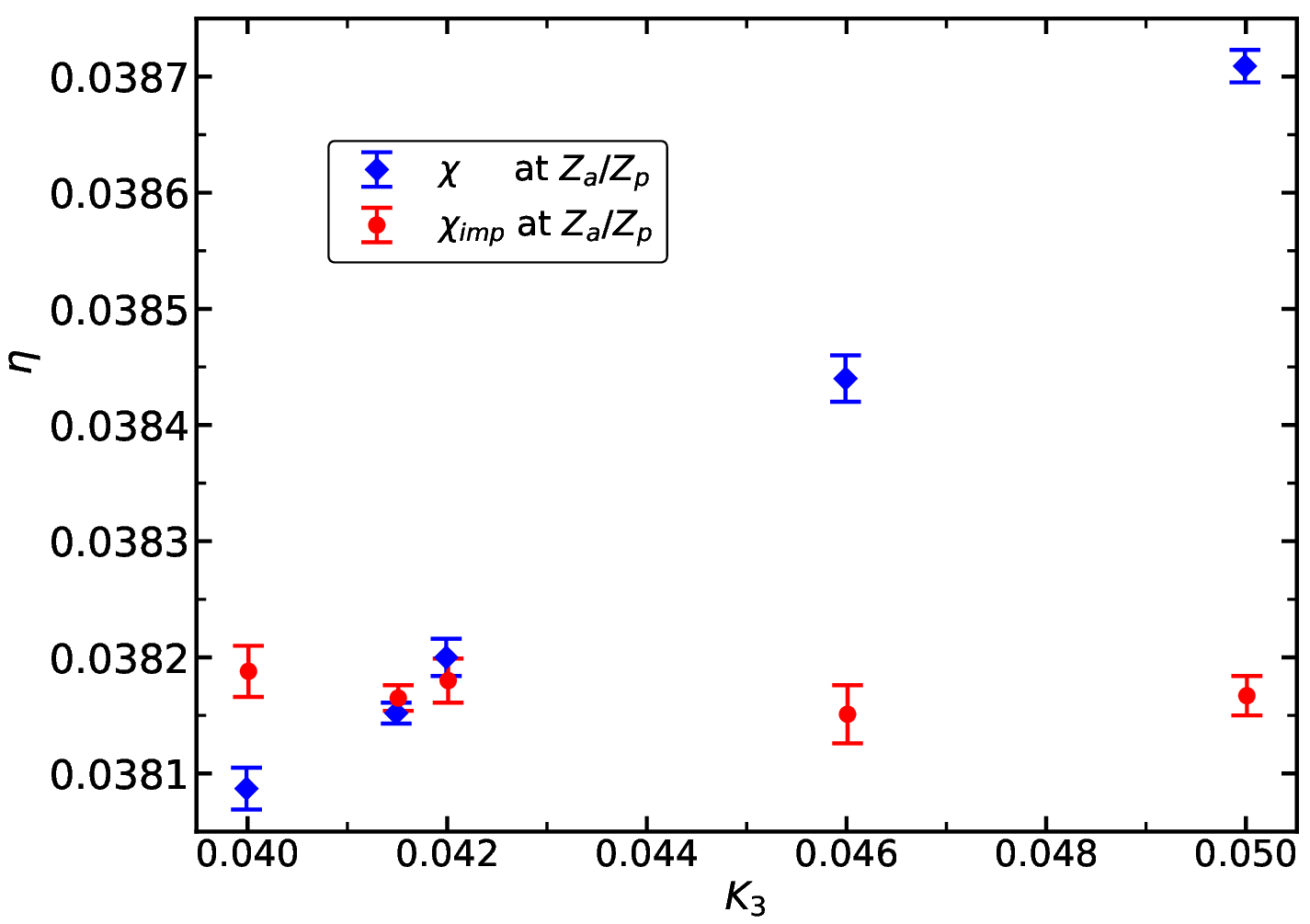}
\caption{\label{chi_imp}
We plot estimates of $\eta$ obtained by using the Ansatz~(\ref{chi_basic})
for the model $D \rightarrow \infty$ and $K_3 > 0$ as a function of $K_3$. 
Throughout, the minimal lattice size $L_{min}=16$ is used. We give results 
obtained by fitting data for $\chi$ as well as $\chi_{imp}$, both at
$Z_a/Z_p=0.32038$.
}
\end{center}
\end{figure}

Furthermore, we consider multiplying $\chi$ by a power of $U_4$, 
such that the correction $c L^{-\omega}$ vanishes for any value of $(D,K_3)$.
We have used $\chi_{imp} = \chi U_4^{p}$,  where $p=-0.98(1)$ for fixing 
$Z_a/Z_p=0.32038$ and $p=-0.45(1)$ for $\xi_{2nd}/L=0.592363$.
In the appendix \ref{ImprovedO}, we discuss how we extract these values 
from data for $q=8$ and $K_3=0$ at $D=0.9$ and $D=1.24$ that were generated for
Ref. \cite{myClock}. Since $p$ is universal, it applies to the model
$D \rightarrow \infty$ and $K_3 > 0$ studied here. 
As check, we compare results for $\eta$ obtained for $\chi$ and 
$\chi_{imp}$ at $Z_a/Z_p=0.32038$. These are obtained by using the Ansatz
\begin{equation}
\label{chi_basic}
 \chi= a L^{2-\eta} + b \;.
\end{equation}
In this Ansatz, we omit leading corrections. Hence we expect a 
dependence of the estimate of $\eta$ on $K_3$ due the leading correction
that depends on $K_3$. Furthermore we omit subleading corrections. They
are at least partially taken into account by the additive parameter $b$.
The results for $L_{min} = 16$ for $K_3>0$ are shown in Fig. \ref{chi_imp}. 
Note that for $K_3=0.0415$ we get an acceptable 
$p$-value starting from $L_{min} = 10$. 
We find that for $\chi_{imp}$ the estimates of $\eta$ obtained for different
values of $K_3$ are consistent, while this is clearly not the case for 
$\chi$. We see
a significant increase of the estimate of $\eta$ with increasing $K_3$.  
For $K_3=0.0415$, the estimates obtained from $\chi_{imp}$ and $\chi$  are 
consistent as we expect for $K_3^*$. Note that for $K_3=0.05$, we get for
example
$\eta=0.038709(14)$, $0.038660(21)$, $0.038631(31)$, and $0.038632(44)$, for 
$L_{min}=16$, $20$, $24$, and $28$, respectively. Furthermore, 
$\chi^2/$DOF $= 1.890$, $1.199$, $0.771$, and $0.900$ for these  minimal
linear lattice sizes that are included in the fit.
One might be tempted to quote $\eta=0.03863(5)$ as the final result.
However the deviation from our final result or the estimate obtained by using 
the conformal bootstrap method \cite{che19} is about 9 times larger as this 
error estimate. This again reminds us that a $\chi^2/$DOF close to one
or an acceptable $p$-value of the fit says little about possible systematic
errors in the parameters of the Ansatz that are due to corrections, 
which are not taken into account.

Our final estimate of $\eta$ is based on fits of $\chi_{imp}$ at 
$\xi_{2nd}/L= 0.592363$ and  $Z_a/Z_p=0.32038$ at $K_3=0.0415$.
Note that $\chi$ at a fixed value of $Z_a/Z_p$ is less affected by 
corrections, while on the other hand, the statistical error of 
$\chi$ at a fixed value of $\xi_{2nd}/L$ is considerably smaller than
that of $\chi$ at a fixed value of $Z_a/Z_p$. Note that this observation is 
consistent with previous work \cite{myClock,HaPiVi}.
Hence it is a good cross check to analyze both these choices.

\begin{figure}
\begin{center}
\includegraphics[width=14.5cm]{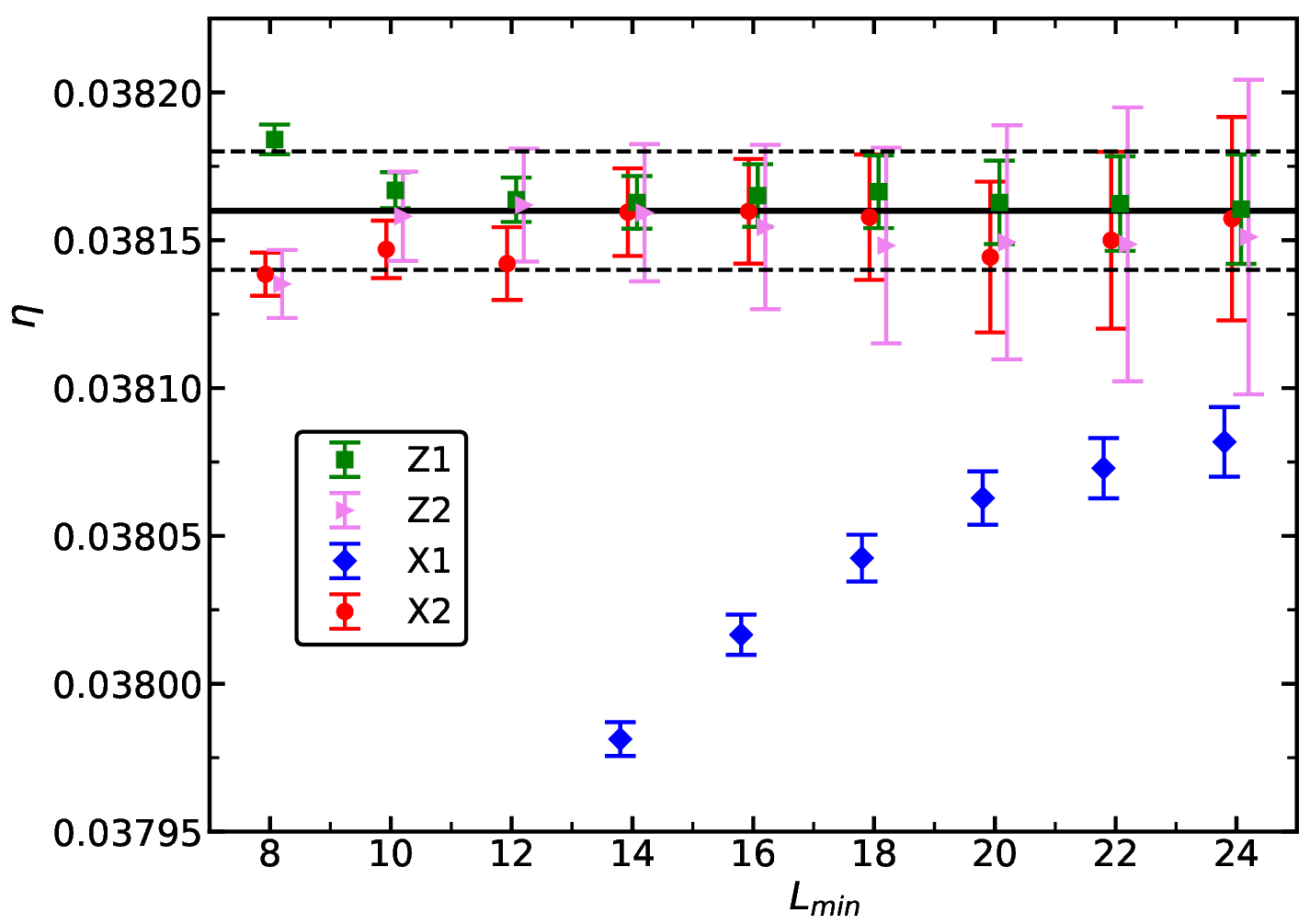}
\caption{\label{final_eta}
We plot estimates of $\eta$ obtained by fitting data for the 
improved magnetic susceptibility $\chi_{imp}$  for $(D,K_3)=(\infty,0.0415)$.
For a discussion of the fits see the text.  The solid line gives our
final estimate, while the dashed lines indicate our error estimate.
Note that the values on the $x$-axis are slightly shifted to reduce overlap
of the symbols.
}
\end{center}
\end{figure}

We used the Ans\"atze~(\ref{chi_basic}) and
\begin{equation}
\label{chi_basic2}
 \chi= a L^{2-\eta} (1+ c L^{-\omega_{NR}})  + b \;,
\end{equation}
where we fix $\omega_{NR} = 2.02548$.
In Fig. \ref{final_eta} the estimates of $\eta$ are plotted versus the
minimal lattice size $L_{min}$ that is taken into account in the fit. In the 
caption of Fig. \ref{final_eta} we denote the data for fixing $\xi_{2nd}/L$  
by X and those 
for fixing $Z_a/Z_p$ by Z. The number that follows, refers to the 
type of the Ansatz. For the Ans\"atze~(\ref{chi_basic}) and (\ref{chi_basic2}) 
there is $1$ or $2$, respectively.  In the case of 
$\chi_{imp}$ at $\xi_{2nd}/L= 0.592363$ using the 
Ansatz~(\ref{chi_basic}) we get an acceptable $p$-value
starting from $L_{min}=20$. 
Still the estimate of $\eta$ is increasing
and it is not consistent with that obtained by the other fits.
For the fit with Ansatz~(\ref{chi_basic2}) we get an acceptable  $p$-value
already for $L_{min}=8$.      
The same holds for $\chi_{imp}$ at $Z_a/Z_p=0.32038$  using the
Ans\"atze~(\ref{chi_basic},\ref{chi_basic2}).
Our final estimate is based on the fits Z1, Z2, and X2, since 
the estimates obtained by using X1 seem to converge much worse than those of
the other three fits. In the case of Ansatz~(\ref{chi_basic2}), 
one should note 
that replacing the numerical value $\omega_{NR} = 2.02548$  by for example 
$2$, the estimates of $\eta$ change only by little. Hence our estimate of 
$\eta$ does not rely critically on the accurate determination of $\omega_{NR}$
by using the CB method. Our final estimate
\begin{equation}
\label{finaleta}
\eta=0.03816(2) \; 
\end{equation}
is chosen such that estimates for some $L_{min}$ for each of Z1, Z2, 
and X2 are contained in the error bar. 
Note that this estimate is consistent with $\eta=0.038176(44)$ obtained 
by using the conformal bootstrap method \cite{che19}. The error is 
considerably reduced compared with our previous estimate 
$\eta=0.03810(8)$ \cite{myClock}.

\begin{figure}
\begin{center}
\includegraphics[width=14.5cm]{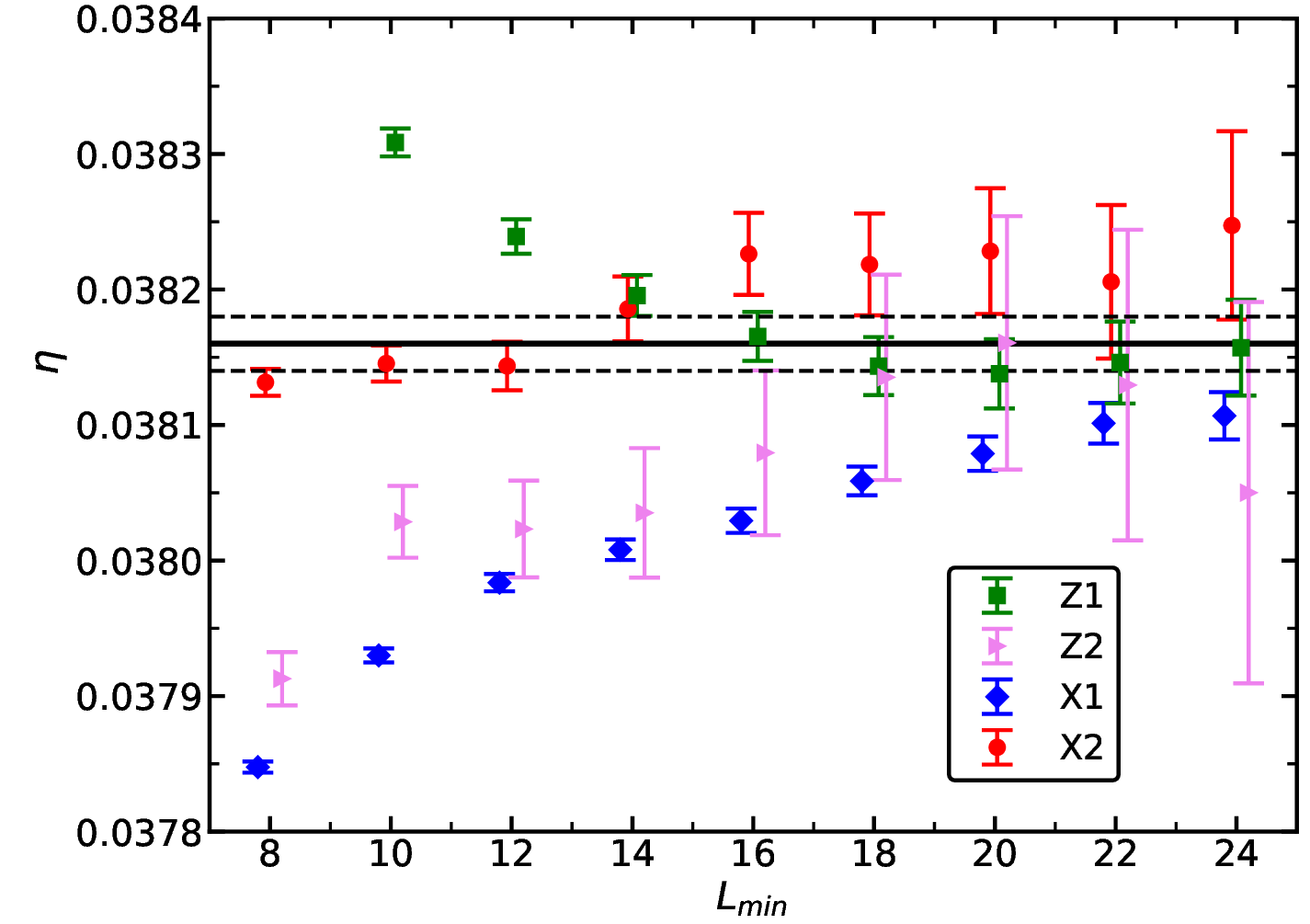}
\caption{\label{final_etaD}
We plot estimates of $\eta$ obtained by fitting data for the
improved magnetic susceptibility $\chi_{imp}$ at $(D,K_3)=(1.06,0)$.
For comparison we give our final estimate obtained by analyzing the 
data for $(D,K_3)=(\infty,0.0415)$ as solid line.
The dashed lines indicate the error of this estimate.
Note that the values on the $x$-axis are slightly shifted to reduce overlap
of the symbols.
}
\end{center}
\end{figure}

As a consistency check, we analyzed our data for $(D,K_3)=(1.06,0)$  in a 
similar fashion. Note that here we have used less CPU time than for 
$(D,K_3)=(\infty,0.0415)$ and the maximal lattice size that we have 
simulated is $L_{max}=100$ instead of  $200$.  Our numerical estimates
are given in Fig. \ref{final_etaD}, which is analogous to Fig. \ref{final_eta}.
The numerical estimates of $\eta$ are fully compatible with the final 
estimate given in Eq.~(\ref{finaleta}). However, one arrives at an error 
bar that is about three times as large.

\subsection{The thermal RG exponent $y_t$}
The thermal RG exponent is extracted from the behavior of the slopes
of dimensionless quantities at criticality.
In general we expect 
\begin{eqnarray}
\label{ytgeneral}
 S &=& a(D,K_3) L^{y_t} \; 
[1 + b(D,K_3) L^{-\omega} +c b^2(D,K_3) L^{-2 \omega } + 
... \nonumber \\
&& + f(D,K_3) L^{-2+\eta}
 + d(D,K_3) L^{-\omega_{NR}}+ ...]  
 + e(D,K_3) L^{-\omega} + ... \;.
\end{eqnarray}
For a derivation of this behavior see for example Sec. III C of Ref.
\cite{myClock}.  The term $f(D,K_3) L^{-2+\eta}$ originates from 
the analytic background of the magnetic susceptibility.
It is present for $U_4$, $U_6$, $\xi_{2nd}/L$, but not for $Z_a/Z_p$. 
Therefore we give preference to $Z_a/Z_p$ estimating $y_t$. 
Note that the term $e(D,K_3) L^{-\omega}$ originates from the 
partial derivative of the dimensionless quantity $R$ with respect to the 
scaling field of the leading correction.  Effectively it amounts to a 
correction with the correction exponent $y_t + \omega \approx 2.28$,
which is only somewhat larger than $\omega_{NR}$.

Using the results of section \ref{dimless}
we can construct linear combinations of dimensionless quantities that do not 
on depend this scaling field. Here we shall analyze the slope of 
$Z_a/Z_p- c_{ Z_a/Z_p, U_4} U_4$, where the numerical value 
of the coefficient is taken from Eq.~(\ref{ratioU}).

\begin{figure}
\begin{center}
\includegraphics[width=14.5cm]{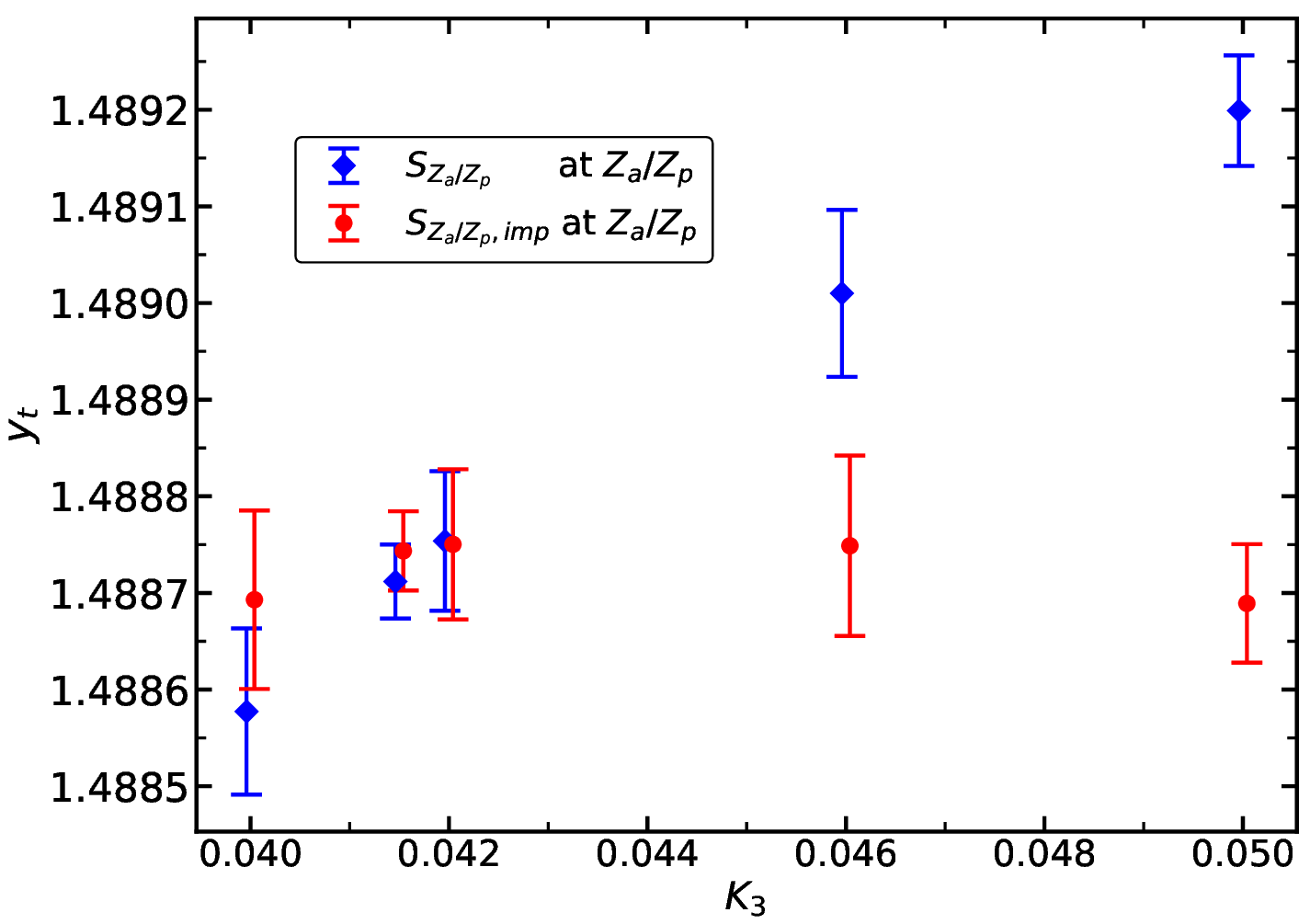}
\caption{\label{slope_vgl}
We plot estimates of $y_t$ obtained by using the Ansatz~(\ref{simple_yt})
for the model $D \rightarrow \infty$ and $K_3 > 0$ as a function of $K_3$.
Throughout, the minimal lattice size $L_{min}=16$ is used. 
We give results obtained
by fitting data for $S_{Z_a/Z_p}$ as well as $S_{Z_a/Z_p,imp}$, both
at $Z_a/Z_p=0.32038$, using the Ansatz~(\ref{simple_yt}).
}
\end{center}
\end{figure}

In a preliminary step, for $D\rightarrow \infty$, we compute an estimate of 
$y_t$ for each value of $K_3$ separately.
To this end, we fit the data by using the Ansatz
\begin{equation}
\label{simple_yt}
 S = a L^{y_t} \; (1 + b L^{-\epsilon})  \;,
\end{equation}
where we take $\epsilon=\omega_{NR} = 2.02548$. Similar to the preliminary 
analysis of the magnetic susceptibility, we compare the estimates obtained
for $S_{Z_a/Z_p,imp}$ and $S_{Z_a/Z_p}$ both at $Z_a/Z_p=0.32038$.  The
improved slope is obtained by using the exponent $p=0.92(4)$.
In Fig. \ref{slope_vgl}
we give results obtained for $L_{min}=16$ for all values of $K_3$ studied here. 
Qualitatively, we see a similar behavior as for the estimates of $\eta$. 
In the case of $S_{Z_a/Z_p}$ we see a dependence of the estimate of $\nu$ on 
$K_3$, while this is not the case for $S_{Z_a/Z_p,imp}$. However this effect 
is less pronounced than in the case of $\eta$, studied above.

\begin{figure}
\begin{center}
\includegraphics[width=14.5cm]{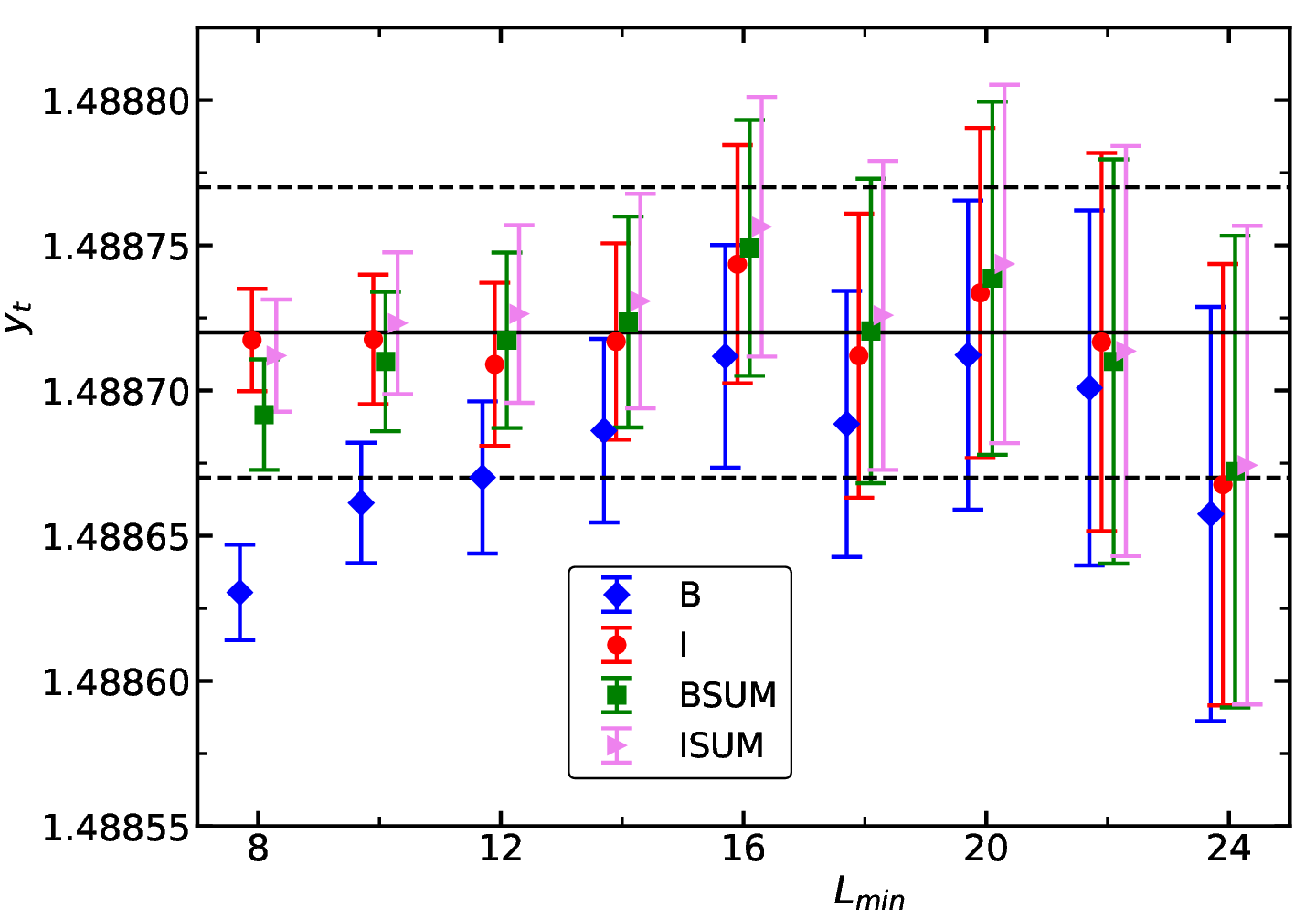}
\caption{\label{yt0415}
We plot estimates of $y_t$ obtained by using the Ansatz~(\ref{simple_yt}) 
versus the minimal lattice size that is included into the fit.
Only data taken at $K_3=0.0415$ are used. Four different slopes are analyzed.
Details and the meaning of B, I, BSUM, and ISUM are discussed in the text. 
The solid line gives our
final estimate, while the dashed lines indicate our error estimate.
Note that the values on the $x$-axis are 
slightly shifted to reduce overlap of the symbols.
}
\end{center}
\end{figure}

Next we focus on $K_3=0.0415$, which is close to $K_3^*$, and where 
we have accumulated the highest statistics and reached the largest 
linear lattice size simulated in this study. Also here, we used 
the Ansatz~(\ref{simple_yt}). In Fig. \ref{yt0415}  we plot estimates 
of $y_t$ as a function of $L_{min}$ obtained for $S_{Z_a/Z_p}$ (B),
$S_{Z_a/Z_p,imp}$ (I), $S_{Z_a/Z_p- c_{ Z_a/Z_p, U_4} U_4}$ (BSUM), 
and $S_{Z_a/Z_p- c_{ Z_a/Z_p, U_4} U_4,imp}$ (ISUM), 
all taken at $Z_a/Z_p=0.32038$. 
The improved slope $S_{Z_a/Z_p- c_{ Z_a/Z_p, U_4} U_4,imp}$ is obtained 
by using the exponent $p=0.22(4)$.
The short names B, I, BSUM, and ISUM are
used in the caption. The results shown in Fig. \ref{yt0415}
are obtained by using $\epsilon=2.02548$ in the Ansatz~(\ref{simple_yt}). 
We have checked that the results change only slightly when using 
$\epsilon=2-\eta$ instead.

These different slopes should be affected by corrections not fully 
taken into account in the Ansatz in different ways: 
\begin{itemize}
\item \noindent B: small corrections $\propto L^{-\omega}$, 
         corrections $\propto L^{-y_t-\omega}$.
\item I: very small corrections $\propto L^{-\omega}$, 
                    corrections $\propto L^{-2+\eta}$, 
                  corrections $\propto L^{-y_t-\omega}$.
\item BSUM: small corrections $\propto L^{-\omega}$, 
             corrections $\propto L^{-2+\eta}$, 
             small corrections $\propto L^{-y_t-\omega}$.
\item ISUM: very small corrections $\propto L^{-\omega}$,
            corrections $\propto L^{-2+\eta}$,
            small corrections $\propto L^{-y_t-\omega}$.
\end{itemize}
Therefore the spread of the results should indicate the size of
the systematic errors.
Assuming that the spectrum of operators looks similar as for the Ising 
universality class, $\omega_i \gtrapprox 3.4$ for corrections not discussed 
above. 

The solid line in Fig. \ref{yt0415} gives our final result for $y_t$ 
and the dashed lines give the error. The error is chosen such that for each
of the four fits there is at least one $L_{min}$ with an acceptable $p$-value
that gives a consistent estimate. In the 
appendix A of Ref. \cite{myCubic2} we discuss in more detail how we arrive 
at final results and their error.
Here we obtain
\begin{equation}
 y_t=1.48872(5)  \;,
\end{equation}
which is consistent but more precise that our previous estimate 
$y_t=1.48879(14)$ \cite{myClock}. 
CB \cite{che19} gives $y_t=3-1.51136(22)=1.48864(22)$. For a more
comprehensive comparison with the literature see Sec. \ref{summary} below.

As check, we analyzed the slopes of $U_4$ and $\xi_{2nd}/L$. We get 
consistent but less accurate results. The same holds for analyzing 
the data for $(D,K_3)=(1.06,0)$.  We abstain from a detailed discussion.

\section{Summary and conclusions}
\label{summary}
We further advance the programme of improved lattice models. The basic idea 
of improved models is that a parameter of the reduced Hamiltonian is 
tuned such that the leading correction to scaling vanishes.
This approach has to be contrasted with efforts to construct 
perfect actions (reduced Hamiltonians) as it was proposed in Ref. 
\cite{Perfect} for example. For a discussion of improved models, 
where one parameter is tuned see for example Sec. 2.3 of the 
review \cite{PeVi02}. The success of improved models relies on 
the fact that the spectrum of correction exponents is sparse at
least for the leading corrections.
The aim of studying improved models is to obtain 
accurate results for universal quantities. Recent applications
to the Ising, $XY$, Heisenberg and $O(4)$-invariant universality classes 
in three dimensions are Refs. \cite{myBlume,myClock,myIco,myLargeN}.
In these cases $\omega \approx 0.8$. For the diluted Ising 
universality class the correction exponent is
$\omega=0.33(3)$ \cite{diluted07}. 
In this case it is virtually mandatory to simulate at the optimal
dilution parameter
to obtain accurate estimates of the critical exponents. In general,
the smaller the correction exponent, the more important is the
improvement to obtain accurate estimates of universal quantities.
A natural question is whether the program can be extended to more 
than one correction.  Recently we have studied the $O(N)$-invariant 
$\phi^4$ model in three dimensions with a perturbation that has only
cubic symmetry for $N=3$ and $4$. Here $\omega_1 =0.763(24)$ and
$\omega_2 =0.082(5)$ for $N=4$ and $\omega_2 = 0.0133(8)$ for $N=3$
\cite{myCubic,myCubic2} for the cubic fixed point. In the case of $N=3$ it is 
plausible that the value of $\omega_1$ is very close to 
$\omega=0.759(2)$ \cite{myIco} for the $O(3)$-invariant fixed point.
Note that the choice of the subscript of $\omega$ follows the literature.
Here it is virtually impossible to obtain accurate results for the critical
exponents of the cubic fixed point without improvement of the reduced 
Hamiltonian. In Refs. \cite{myCubic,myCubic2} we eliminate the two
leading corrections by tuning two parameters of the reduced Hamiltonian.

In Ref. \cite{myIso} we study the Blume-Capel model on the simple cubic 
lattice with next-to-next-to-nearest neighbor interactions in addition
to nearest neighbor interactions. Essentially by tuning the ratio of the 
corresponding couplings $K_3/K_1$, we eliminate the leading violation
of spatial rotational invariance, while 
the leading correction is eliminated by additionally tuning the dynamic 
dilution, or crystal field,
parameter $D$. Highly accurate estimates of critical 
exponents for the Ising universality class were obtained.

In the present work we investigate whether this approach can be applied 
successfully in the case of the XY universality class. We build on 
a preliminary study presented in the appendix of Ref. \cite{myIso}. 
We find that in a rigorous sense this is not the case. However for
$D \rightarrow \infty$, leading corrections are eliminated for 
$(K_3/K_1)^* =0.1119(5)$, where the spatial unisotropy is reduced by a factor
of about 9 compared with the model at $K_3=0$.
The price to pay for $K_3>0$ is the excess CPU time needed compared with the 
$q$-state clock model with nearest neighbor couplings only. 
The $q$-state clock  model can be simulated
by using the single cluster algorithm \cite{Wolff} only. The main task in the 
algorithm is to check, whether a link to the nearest neighbor, and here
in addition the next-to-next-to-nearest neighbor, is 
frozen or not. Since there are six nearest neighbors and eight 
next-to-next-to-nearest neighbors, we naively expect that the 
CPU time needed increases by the factor $(6+8)/6=2.3...$. Comparing 
our own implementations of the update, we see a factor slightly 
smaller than two.

At $K_3=0.0415$, which is close to $K_3^*$, we performed high statistics
simulations on lattices with a linear lattice size up to $L=200$. A 
FSS analysis of the data provided us with accurate estimates
of the critical exponents $\eta$ and $y_t =1/\nu$. 
In table \ref{summary_tab}
our results are compared with ones given in the literature. These were 
obtained by using various theoretical methods and experiments. In particular
in the case of field theoretic methods it is virtually impossible 
to give a complete overview on the vast literature. 

The $\epsilon$-expansion has been introduced in Ref. \cite{WiFi72}, where
the leading order result is given. The expansion parameter is
$\epsilon=4-d$, where $d$ is the dimension of the system. Over the 
years, coefficients for higher orders were computed. For an account
see table 3 of Ref. \cite{He23}. It turns our that the series is divergent
and in order to obtain a numerical result for the critical exponents 
in three dimensions resummation of the series is required.
Various methods have been applied to this end.
In general the systematic error is hard to estimate and hence the 
error that is quoted depends on the judgment of the authors. In table
\ref{summary_tab} we give a small selection of results. The precision
is clearly less than that obtained here or by using CB. 
The same holds for the results of the expansion in three dimensions fixed 
\cite{GuZi98} and FRG \cite{DePo20}.
For more field theoretical results see table 19 of Ref. \cite{PeVi02}.

Lattice models can be studied by various methods. Precise estimates
of critical exponents of three-dimensional systems are obtained 
by analyzing high temperature series expansions or Monte Carlo
simulations. In Ref. \cite{BuCo97} the high temperature
series of the susceptibility and the correlation length were computed 
up to the order $\beta^{21}$, where $\beta$ is the inverse temperature,
for $O(N)$-invariant nonlinear $\sigma$ 
models on the simple cubic and the body centered cubic lattice. Numerical
results for the exponents $\gamma$ and $\nu$ are given in table II 
of Ref. \cite{BuCo97}. In table \ref{summary_tab} we give estimates for
$N=2$, where we compute $\eta=2 -\gamma/\nu$. 
We abstain from quoting an error for $\eta$, since it is 
unclear how $\gamma$ and $\nu$ are correlated. For previous work 
see Ref. \cite{BuCo97}.

An early Monte Carlo study of the XY model on the simple cubic 
lattice is Ref. \cite{Swendsen83}. Using the Monte Carlo renormalization
group (MCRG), estimates of the RG exponents $y_h$ and $y_t$ were computed.
However no final estimate is quoted. It is interesting to note 
that in addition to the standard XY model an 
improved model is simulated. In addition to the nearest neighbor 
coupling a second term, Eq.~(11) of Ref. \cite{Swendsen83}, is added in 
the Hamiltonian. For a discussion see sec. IV. of Ref. \cite{Swendsen83}. 
The advent of the cluster 
algorithm \cite{SwWa87,Wolff} was a milestone in the simulation of 
$O(N)$-invariant lattice
models. This can be clearly seen when comparing the results of 
Ref. \cite{Janke90} with \cite{LiTe89}. In Ref. \cite{Janke90}, 
where the single cluster algorithm has been used, much more precise 
estimates are obtained than in Ref. \cite{LiTe89}, where a local 
algorithm had been employed.
In Refs. \cite{HaTo99,XY1,XY2,myClock} improved models were simulated by 
using cluster algorithms.
In Refs. \cite{XY1,XY2} the most accurate estimates of the 
critical exponents are actually obtained from the analysis of 
high temperature
series. The analysis is biased by using the estimates of $\lambda^*$, $D^*$
and $\beta_c$ obtained from Monte Carlo simulations in combination with FSS. 
Note that $\lambda$ is the parameter of the $\phi^4$ model that has been 
studied in Refs. \cite{XY1,XY2}. In Ref. \cite{Xu19} two classical
models and a quantum model are simulated using worm-type algorithms.

Let us turn to experiments.
In Ref. \cite{SiAh84} the $\lambda$-transition of $^4$He has been 
studied. The exponent $\nu$ has been determined from the behavior of 
the superfluid fraction, which is obtained from the second sound 
velocity. This estimate is fully consistent with recent theoretical 
results. In Refs. \cite{Lipa96,Lipa00,Lipa03} the specific heat 
in the neighborhood of the $\lambda$-transition
has been measured on a Spacelab mission, avoiding gravitational
rounding of the transition. The final analysis of the data provided
the estimate $\alpha =-0.0127(3)$, which is converted in table 
\ref{summary_tab} by using the hyperscaling relation
$\nu = (2 -\alpha)/d$. The deviation of this result from recent 
theoretical estimates is 8 times larger than the error that 
is quoted. Phase transitions in the XY universality class are 
also expected for other materials. For example magnets with 
an easy axis. A more recent experiment on CsMnF$_3$,
where the specific heat (C) and the thermal diffusivity (D) have 
been measured, is reported in Ref. \cite{OlSaBu14}. In addition, 
data from Ref. \cite{Oletal12} for the specific heat of SmMnO$_3$ 
have been analyzed in \cite{OlSaBu14}. These results are quite accurate 
and more or less agree with the theoretical one. 
For more experimental results see table 20 of Ref. \cite{PeVi02}.

\begin{table}
\caption{\sl \label{summary_tab}
We summarize estimates of critical exponents of the three-dimensional 
XY universality class. MC: Monte Carlo simulation of  lattice models.
HT: Analysis of high temperature series of lattice models. FRG:
functional renormalization group. CB: conformal bootstrap.
 $^*$ means that the exponent is not
directly given in the cited paper. For a discussion see the text.
}
\begin{center}
\begin{tabular}{ccclll}
\hline
\mc{1}{c}{method} &
\mc{1}{c}{year} &
\mc{1}{c}{Ref.} &
\mc{1}{c}{$\nu$} &
\mc{1}{c}{$\eta$} & 
\mc{1}{c}{$\omega$} \\
\hline
3D-exp     &1998 &\cite{GuZi98} & 0.6703(15) & 0.0354(25) &  0.789(11) \\
\hline
$\epsilon$ 5-loop &1998 & \cite{GuZi98} &0.6680(35)  & 0.0380(50) & 0.802(18) \\ 
$\epsilon$ 6-loop &2017 & \cite{KoPa17} &0.6690(10)  & 0.0380(6)  & 0.804(3) \\ 
$\epsilon$ 7-loop &2021 & \cite{Sch18,Sha21}&0.67076(38) & 0.03810(56)& 0.789(13) \\
\hline
FRG   & 2020 &  \cite{DePo20} & 0.6716(6) & 0.0380(13) & 0.791(8) \\
\hline 
CB    & 2016 & \cite{Kos16}  & 0.6719(12)    & 0.0385(7)    & 0.811(19) \\
CB    & 2019 & \cite{che19}  & 0.671754(100) & 0.038176(44) & 0.794(8)\\
\hline
HT, sc  & 1997 &\cite{BuCo97}   &
0.677(3) &  0.0399$^*$  &     -   \\
HT, bcc  & 1997 &\cite{BuCo97}   &
0.674(2) & 0.0386$^*$  &     -   \\
\hline
 MC   & 1989& \cite{LiTe89} &0.67(2)     & 0.075$^*$      &   - \\
 MC   & 1990& \cite{Janke90}&0.670(2)    & 0.036(14)    &   - \\
 MC   & 1999& \cite{HaTo99} &0.6723(3)[8]& 0.0381(2)[2] & 0.79(2) \\
 MC+HT& 2001& \cite{XY1}& 0.67155(27)  & 0.0380(4)  &  - \\
 MC+HT& 2006& \cite{XY2}& 0.6717(1)    & 0.0381(2)  & 0.785(20) \\
 MC   & 2019 & \cite{Xu19}&0.67183(18)  &0.03853(48) & 0.77(13) \\
 MC  & 2019 & \cite{myClock}& 0.67169(7) & 0.03810(8) & 0.789(4) \\
 MC  & 2025 & this work & 0.671718(23) & 0.03816(2) & - \\
\hline
$^4$He & 1984 & \cite{SiAh84} & 0.6717(4) & - & - \\
$^4$He & 2003 & \cite{Lipa96,Lipa00,Lipa03} & 0.6709(1) & - & - \\
SmMnO$_3$,C& 2012 & \cite{Oletal12,OlSaBu14}    & 0.6710(3)& - & - \\
CsMnF$_3$,D& 2014 & \cite{OlSaBu14}             & 0.6710(7)& - & - \\
CsMnF$_3$,C& 2014 & \cite{OlSaBu14}             & 0.6720(13)& - & - \\
\hline
\end{tabular}
\end{center}
\end{table}

The estimates obtained here could be improved simply by spending more CPU time
to get better statistics and larger lattice sizes.
Performing high temperature series expansions for the $q$-state clock model
with nearest and next-to-next-to-nearest neighbor couplings might also
allow to further improve the accuracy of critical exponents. The estimates
of $K_3^*$ and $K_{1,c}$ obtained here could be used to bias the analysis 
of the series.

The model studied here could be used to study universal amplitude ratios,
critical dynamical behavior, and the physics of boundaries and thin films in 
the neighborhood of the critical temperature.

\section{Acknowledgement}
I like to thank Slava Rychkov for pointing my attention to 
refs. \cite{O2corrections,He23} and Junyu Liu and David Simmons-Duffin for 
communicating the numerical value of $\omega_{NR}$.
This work was supported by the DFG under the grants No HA 3150/5-3
and HA 3150/5-4.

\appendix
\section{Improved observables}
\label{ImprovedO}
We eliminate leading corrections in the magnetic susceptibility
and the slopes of dimensionless quantities by multiplying with a
power $p$ of $U_4$.  This idea goes back to Refs. \cite{XY2,diluted07}.

Simulating improved models, these quantities are useful, since 
$D^*$ or $K_3^*$ are only known approximately.

Let us discuss the method at the example of a slope, say at a fixed 
value of $Z_a/Z_p$:
\begin{eqnarray}
\label{simpleAnsatz}
S &=& a L^{y_t} (1 +c_s L^{-\omega}) \nonumber \\
U_4 &=& U_4^*  (1 + c_U L^{-\omega}) \;,
\end{eqnarray} 
where we only give the leading correction.
We intent to construct 
\begin{equation}
 S_{imp} = S U_4^p  
         = a U_4^{*,p} L^{y_t} (1+c_s L^{-\omega}) (1 + p c_U L^{-\omega} + ...) 
         = a U_4^{*,p} L^{y_t} (1+[c_s+ p c_U] L^{-\omega} + ...) 
\end{equation}
such that leading corrections vanish.
Hence $p=-c_s/c_U$. In order to get estimates of $c_s$ and $c_U$, one
could perform simulations at values of $D$ or $K_3$, where leading corrections
to scaling have a large amplitude and then fit the data by using 
Eqs.~(\ref{simpleAnsatz}) as Ans\"atze. However it is favorable to
eliminate RG exponents first. To this end we consider a pair 
of parameters, where the leading corrections have opposite sign and 
roughly the same absolute value. To this end, we take data from 
simulations performed for \cite{myClock} for the $(8+1)$-state clock  
model at $D=0.9$ and $1.24$. Computing the ratio
\begin{equation}
r_S(L) =\frac{S(L,D=0.9)}{S(L,D=1.24)}
\end{equation}
we eliminate $L^{y_t}$.  Analogously $r_U(L) =U_4(L,D=0.9)/U_4(L,D=1.24)$,
We perform fits by using the Ansatz
\begin{equation}
r_S(L) = \bar{a} r_U(L)^{-p} \;,
\end{equation}
where $\bar{a}$ and $p$ are the free parameters of the fit. Note that
in this Ansatz $\omega$ does not appear. In the fit we ignore the 
statistical error of $r_U(L)$, since its relative error is much smaller
than that of $r_S(L)$. For $D=0.9$ and $D=1.24$, we had simulated lattices
of a linear size up to $L=48$. The estimates of $p$ that we obtain are given 
in the main text.

\def\refname{}


\begin{thebibliography}{99}
\bibitem{WiKo}
K. G. Wilson and J. Kogut,
{\sl The renormalization group and the $\epsilon$-expansion},
Phys.\ Rep.\ C {\bf 12}, 75 (1974).

\bibitem{Fisher74}
M. E. Fisher,
{\sl The renormalization group in the theory of critical behavior},
Rev.\ Mod.\ Phys.\ {\bf 46}, 597 (1974), Erratum: 
Rev.\ Mod.\ Phys.\ {\bf 47}, 543 (1975).

\bibitem{Cardy}
John Cardy, {\sl Scaling and Renormalization in Statistical Physics},
Series: Cambridge Lecture Notes in Physics (No. 5)
(Cambridge University Press, Cambridge, 1996)

\bibitem{Fisher98}
M. E. Fisher,
{\sl Renormalization group theory: Its basis and formulation in statistical physics},
Rev.\ Mod.\ Phys.\ {\bf 70}, 653 (1998).

\bibitem{PeVi02}
A. Pelissetto and E. Vicari,
{\sl Critical Phenomena and Renormalization-Group Theory},
[arXiv:cond-mat/0012164],
Phys.\ Rept.\ {\bf 368}, 549 (2002).

\bibitem{SwWa87}
Robert H. Swendsen and Jian-Sheng Wang,
{\sl Nonuniversal critical dynamics in Monte Carlo simulations},
Phys.\ Rev.\ Lett.\ {\bf 58}, 86 (1987). 

\bibitem{Wolff}
U. Wolff,
{\sl Collective Monte Carlo Updating for Spin Systems},
Phys.\ Rev.\ Lett.\ {\bf 62}, 361 (1989).

\bibitem{myIso}
Martin Hasenbusch,
{\sl Restoring isotropy in a three-dimensional lattice model: 
The Ising universality class},
[arXiv:2105.09781], Phys.\ Rev.\ B {\bf 104}, 014426 (2021).

\bibitem{Kos16}
F. Kos, D. Poland, D. Simmons-Duffin, and A. Vichi,
{Precision Islands in the Ising and $O(N)$ Models},
[arXiv:1603.04436], J.\ High Energ.\ Phys.\ 08 (2016), 36.

\bibitem{SD16}
 D. Simmons-Duffin,
{\sl The Lightcone Bootstrap and the Spectrum of the 3d Ising CFT},
[arXiv:1612.08471],   J.\ High Energ.\ Phys.\ 03 (2017), 86.

\bibitem{Re22}
M. Reehorst, {\sl Rigorous bounds on irrelevant operators in the 3d Ising
model CFT}, [arXiv:2111.12093]  J.\ High Energ.\ Phys.\ 09 (2022), 177.

\bibitem{CB_Ising_2024}
Cyuan-Han Chang, Vasiliy Dommes, Rajeev S. Erramilli, Alexandre Homrich,
Petr Kravchuk, Aike Liu, Matthew S. Mitchell, David Poland,
David Simmons-Duffin,
{\sl Bootstrapping the 3d Ising Stress Tensor},
[arXiv:2411.15300], J.\ High Energ.\ Phys.\  03 (2025), 136.

\bibitem{myClock}
M. Hasenbusch,
{\sl Monte Carlo study of an improved clock model in three dimensions},
[arXiv:1910.05916],
Phys.\ Rev.\ B {\bf 100}, 224517 (2019).

\bibitem{che19}
S. M. Chester, W. Landry, J. Liu, D. Poland, D. Simmons-Duffin, N. Su, 
and A. Vichi,
{\sl Carving out OPE space and precise $O(2)$ model critical exponents},
[arXiv:1912.03324], J.\ High Energ.\ Phys.\  06 (2020), 142.

\bibitem{PaRyVi18}
D. Poland, S. Rychkov, and A. Vichi,
{\sl The Conformal Bootstrap: Theory, Numerical Techniques, and Applications},
[arXiv:1805.04405], Rev.\ Mod.\ Phys.\ {\bf 91}, 15002 (2019).
%

\bibitem{RySu23}
Slava Rychkov and Ning Su,
{\sl New Developments in the Numerical Conformal Bootstrap}, 
[arXiv:2311.15844], Rev.\ Mod.\ Phys.\ {\bf 96}, 045004 (2024).

\bibitem{Lipa96}
J. A. Lipa, D. R. Swanson, J. A. Nissen, T. C. P. Chui, and U. E.
Israelsson,
{\sl  Heat Capacity and Thermal Relaxation of Bulk Helium very near the
Lambda Point}, Phys.\ Rev.\ Lett.\ {\bf 76}, 944  (1996).

\bibitem{Lipa00}
J. A. Lipa, D. R. Swanson, J. A. Nissen, Z. K. Geng, P. R. Williamson,
D. A. Stricker, T. C. P. Chui, U. E. Israelsson, and M. Larson,
{\sl Specific Heat of Helium Confined to a 57- $\mu$m Planar Geometry},
Phys.\ Rev.\ Lett.\  {\bf 84}, 4894 (2000).

\bibitem{Lipa03}
J. A. Lipa, J. A. Nissen, D. A. Stricker, D. R. Swanson and T. C. P. Chui,
{ \sl
Specific heat of liquid helium in zero gravity very near the $\lambda$-point},
[arXiv:cond-mat/0310163],
Phys.\ Rev.\ B {\bf 68}, 174518 (2003).

\bibitem{GuZi98}
R. Guida and J. Zinn-Justin,
{\sl Critical exponents of the N vector model},
[arXiv:cond-mat/9803240], J.\ Phys.\ A {\bf 31}, 8103 (1998).

\bibitem{XY2}
M. Campostrini, M. Hasenbusch, A. Pelissetto, and E. Vicari,
{\sl The critical exponents of the superfluid transition in He4},
[cond-mat/0605083], published as
{\sl Theoretical estimates of the critical exponents of the superfluid
transition in He4 by lattice methods}, Phys.\ Rev.\ B {\bf 74}, 144506 (2006).\\

\bibitem{Xu19}
Wanwan Xu, Yanan Sun, Jian-Ping Lv, and Youjin Deng,
{\sl High-precision Monte Carlo study of several models in
the three-dimensional $U(1)$ universality class},
[arXiv:1908.10990], Phys.\ Rev.\ B {\bf 100}, 064525 (2019).

\bibitem{LiTe89}
Ying-Hong Li and S. Teitel,
{\sl Finite-size scaling study of the three-dimensional classical XY model}, 
Phys.\ Rev.\ B {\bf 40}, 9122 (1989).

\bibitem{KoOk14}
Yukihiro Komura and Yutaka Okabe,
{\sl CUDA programs for GPU computing of Swendsen-Wang multi-cluster spin ﬂip algorithm:
2D and 3D Ising, Potts, and XY models},
[arXiv:1403.7560], Computer Physics Communications 185 (2014) 1038.

\bibitem{XY1}
M. Campostrini, M. Hasenbusch, A. Pelissetto, P. Rossi, and
E. Vicari,
{\sl Critical behavior of the three-dimensional XY universality class},
[cond-mat/0010360], Phys.\ Rev.\ B {\bf 63}, 214503  (2001). \\

\bibitem{He23}
Johan Henriksson,
{\sl The critical O(N) CFT: Methods and conformal data},
[arXiv:2201.09520], Physics Reports {\bf 1002}, 1 (2023).

\bibitem{Nelson_danger}
David R. Nelson,
{\sl Coexistence-curve singularities in isotropic ferromagnets},
Phys.\ Rev.\ B {\bf 13}, 2222 (1976).

\bibitem{AmPe82}
Daniel J. Amit and Luca Peliti, 
{\sl On dangerous irrelevant operators},
Annals of Physics {\bf 140}, 207 (1982).

\bibitem{Alvarez}
L. Alvarez-Gaume, O. Loukas, D. Orlando, and S. Reffert,
{\sl Compensating strong coupling with large charge},
[arXiv:1610.04495], J.\ High Energ.\ Phys.\  04 (2017), 059.

\bibitem{Debasish}
D. Banerjee, S. Chandrasekharan, and D. Orlando,
{\sl Conformal dimensions via large charge expansion},
[arXiv:1707.00711], Phys.\ Rev.\ Lett.\ {\bf 120}, 061603 (2018).

\bibitem{Cuomo23}
Gabriel Cuomo, J. M. Viana Parente Lopes, Jos\'e Matos, J\'ulio Oliveira,
and Jo\~ao Penedones,
{\sl Numerical tests of the large charge expansion},
[arXiv:2305.00499], J.\ High Energ.\ Phys.\  05 (2024), 161.

\bibitem{O234}
M. Hasenbusch and E. Vicari,
{\sl Anisotropic perturbations in three-dimensional O(N)-symmetric vector
 models},
[arXiv:1108.0491], Phys.\ Rev.\ B {\bf 84}, 125136 (2011).

\bibitem{Hove03}
J. Hove and A. Sudb\o,
{\sl Criticality versus $q$ in the $(2+1)$-dimensional $Z_q$ clock model},
[arXiv:cond-mat/0301499],
Phys.\ Rev.\ E {\bf 68}, 046107 (2003).

\bibitem{Shao19}
H. Shao, W. Guo, and A. W. Sandvik,
{\sl Monte Carlo Renormalization Flows in the Space of Relevant and Irrelevant Operators:
Application to Three-Dimensional Clock Models},
[arXiv:1905.13640], Phys.\ Rev.\ Lett.\ {\bf 124}, 080602 (2020).

\bibitem{fuzzy}
Wei Zhu, Chao Han, Emilie Huffman, Johannes S. Hofmann, and Yin-Chen He, 
{\sl Uncovering conformal symmetry in the 3D Ising transition: 
State-operator correspondence from a fuzzy sphere regularization}, 
[arXiv:2210.13482], 
Phys.\ Rev.\ X {\bf 13}, 021009 (2023).
%

\bibitem{DePo20}
G. De Polsi, I. Balog, M. Tissier, and N. Wschebor,
{\sl Precision calculation of critical exponents in the $O(N)$ universality
classes with the nonperturbative renormalization group},
[arXiv:2001.07525], Phys.\ Rev.\ E {\bf 101}, 042113 (2020).

\bibitem{O2corrections}
Junyu Liu, David Meltzer, David Poland, David Simmons-Duffin,
{\sl The Lorentzian inversion formula and the spectrum of the 3d O(2) CFT}
[arXiv:2007.07914], J.\ High Energ.\ Phys.\  09 (2020), 115.
%

\bibitem{private}
Junyu Liu and David Simmons-Duffin, private communication.

\bibitem{Ma17}
A. N. Manashov, E. D. Skvortsov and M. Strohmaier,
{\sl Higher spin currents in the critical
$O(N)$ vector model at $1/N^2$}, [arXiv:1706.09256],
J.\ High Energ.\ Phys.\  08 (2017), 106.

\bibitem{De98}
S. E. Derkachov, J. A. Gracey and A. N. Manashov,
{\sl Four loop anomalous dimensions of gradient
operators in $\phi^4$ theory}, [arXiv:hep-ph/9705268]
Eur.\ Phys.\ J.\ C {\bf 2}, 569 (1998).

\bibitem{XYamp}
M. Hasenbusch, 
{\sl A Monte Carlo study of the three-dimensional XY universality class: 
Universal amplitude ratios}, 
[arXiv:0810.2716], J. Stat. Mech. (2008) P12006.

\bibitem{MyVar}
M. Hasenbusch, 
{\sl Variance-reduced estimator of the connected two-point function
in the presence of a broken $\mathbb{Z}_2$-symmetry},
[arXiv:1512.02491],  Phys.\ Rev.\ E {\bf 93}, 032140 (2016).

\bibitem{PaKa05}
M. Patra and M. Karttunen,
{\sl Stencils with Isotropic Discretization Error
for Differential Operators},
Numerical Methods for Partial Differential Equations {\bf 22}, 936 (2006).

\bibitem{HaPiVi}
M. Hasenbusch, K. Pinn, and S. Vinti,
{\sl Critical Exponents of the 3D Ising Universality Class From Finite Size 
Scaling With Standard and Improved Actions},
[arXiv:hep-lat/9806012], Phys.\ Rev.\ B {\bf 59}, 11471 (1999).

\bibitem{VignaWWW}
\verb+ https://prng.di.unimi.it/ +

\bibitem{ViBl18}
D. Blackman and S. Vigna,
{\sl Scrambled Linear Pseudorandom Number Generators},
[arXiv:1805.01407], ACM Trans.\ Math.\ Softw.\ {\bf 47}, 36  (2021).

\bibitem{ONeill_minimal}
\verb+ https://www.pcg-random.org/posts/does-it-beat-the-minimal-standard.html+

\bibitem{KISS_wiki}
\verb+ https://de.wikipedia.org/wiki/KISS_(Zufallszahlengenerator) +

\bibitem{twister}
M. Saito and M. Matsumoto,
``SIMD-oriented Fast Mersenne Twister:
a 128-bit Pseudorandom Number Generator'',
in
{\sl Monte Carlo and Quasi-Monte Carlo Methods 2006},
edited by A. Keller, S. Heinrich, H. Niederreiter, (Springer, 2008);
M. Saito, Masters thesis, Math. Dept., Graduate School of science,
Hiroshima University, 2007.
The source code of the program is provided at \\
\verb+http://www.math.sci.hiroshima-u.ac.jp/~m-mat/MT/SFMT/index.html+

\bibitem{myCubic2}
M. Hasenbusch,
{\sl $\phi^4$ lattice model with cubic symmetry in three dimensions: 
Renormalization group flow and first-order phase transitions},
[arXiv:2307.05165], Phys.\ Rev.\ B {\bf 109}, 054420 (2024).

\bibitem{pythonSciPy}
P. Virtanen, R. Gommers, T. E. Oliphant et al.,
{\sl SciPy 1.0--Fundamental Algorithms for Scientific Computing in Python},
[arXiv:1907.10121], Nature Methods {\bf 17}, 261 (2020).

\bibitem{plotting}
J. D. Hunter, {\sl "Matplotlib: A 2D Graphics Environment},
Computing in Science \& Engineering {\bf 9}, 90 (2007).

\bibitem{Perfect}
P. Hasenfratz and F. Niedermayer, 
{\sl Perfect lattice action for asymptotically free theories},
[arXiv:hep-lat/9308004], Nucl.\ Phys.\ B {\bf 414}, 785 (1994).

\bibitem{myBlume}
M. Hasenbusch,
{\sl Finite size scaling study of lattice models in the three-dimensional 
Ising universality class},   [arXiv:1004.4486],
Phys.\ Rev.\ B {\bf 82}, 174433 (2010). 

\bibitem{myIco}
M. Hasenbusch,
{\sl Monte Carlo study of a generalized icosahedral model on the simple cubic
lattice}, [arXiv:2005.04448],  Phys.\ Rev.\ B {\bf 102}, 024406 (2020).

\bibitem{myLargeN}
M. Hasenbusch, {\sl Three-dimensional O$(N)$-invariant models 
at criticality for $N \ge 4$},
[arXiv:2112.03783], Phys.\ Rev.\ B {\bf 105}, 054428 (2022).

\bibitem{diluted07}
Martin Hasenbusch, Francesco Parisen Toldin, Andrea Pelissetto, and
Ettore Vicari,
{\sl Universality class of 3D site-diluted and bond-diluted Ising systems},
[arXiv:cond-mat/0611707], J.\ Stat.\ Mech.\ (2007) P02016.
%

\bibitem{myCubic}
M. Hasenbusch,
{\sl Cubic fixed point in three dimensions: Monte Carlo simulations of the 
model on the simple cubic lattice}, [arXiv:2211.16170],
Phys.\ Rev.\ B {\bf 107}, 024409  (2023).

\bibitem{WiFi72}
 K. G. Wilson and M. E. Fisher, {\sl Critical exponents in 3.99 dimensions}, 
Phys.\ Rev.\ Lett.\ {\bf 28}, 240 (1972).

\bibitem{BuCo97}
P. Butera and M. Comi, 
{\sl 
N-vector spin models on the sc and the bcc lattices: a study of the 
critical behavior
of the susceptibility and of the correlation length by high temperature 
series extended to order $\beta^{21}$
},
[arXiv:hep-lat/9703018],
Phys.\ Rev.\ B {\bf 56}, 8212 (1997).


\bibitem{Swendsen83}
Robert H. Swendsen,
{\sl Monte Carlo renormalization-group study of the $d=3$ planar model},
Phys.\ Rev.\ B {\bf 27}, 391 (1983).

\bibitem{Janke90}
W. Janke, 
{\sl Test of single cluster update for the three-dimensional XY model},
Phys.\ Lett.\ A {\bf 148}, 306 (1990).

\bibitem{HaTo99}
M. Hasenbusch  and T. T\"or\"ok,
{\sl High precision Monte Carlo study of the 3D XY-universality class},
[arXiv:cond-mat/9904408], J. Phys. A: Math. Gen. {\bf 32}, 6361 (1999).

\bibitem{SiAh84}
A. Singasaas and G. Ahlers, 
{\sl Universality of static properties near the superfluid transition in 
$^4$ He},
Phys.\ Rev.\ B {\bf 30}, 5103 (1984).

\bibitem{Oletal12}
A. Oleaga, A. Salazar, D. Prabhakaran, J.G. Cheng, and J.S. Zhou,
{\sl Critical behavior of the paramagnetic to antiferromagnetic 
transition in orthorhombic and hexagonal phases of RMnO3
(R=Sm, Tb, Dy, Ho, Er, Tm, Yb, Lu, Y)}
Phys.\ Rev.\ B {\bf 85}, 184425 (2012).

\bibitem{OlSaBu14}
A. Oleaga, A. Salazar, and Yu M. Bunkov,
{\sl 3D-XY critical behavior of CsMnF$_3$ from static and
dynamic thermal properties},
J.\ Phys.:\ Condens.\ Matter {\bf 26}, 096001 (2014).

\bibitem{KoPa17}
M. V. Kompaniets and E. Panzer,
{\sl Minimally subtracted six-loop renormalization of
$\phi^4$-symmetric theory and critical exponents},
[arXiv:1705.06483], Phys.\ Rev.\  D {\bf 96}, 036016 (2017).

\bibitem{Sch18}
O. Schnetz, Phys.\ Rev.\ D {\bf 97}, 085018 (2018); Maple package
HyperlogProcedrues, which is available on the Oliver Schnetz's
homepage: \verb+https://www.math.fau.de/person/oliver-schnetz/+

\bibitem{Sha21}
Abouzeid M. Shalaby,
{\sl Critical exponents of the O(N)-symmetric $\phi^4$ model from the 
$\epsilon^7$ hypergeometric-Meijer resummation}, [arXiv:2005.12714],
Eur.\ Phys.\ J.\ C {\bf 81}, 87 (2021).

\end{thebibliography}
\end{document}